\documentclass[12pt]{article}

\usepackage{amsfonts,ulem}
\usepackage{amsmath}
\usepackage{amssymb,authblk}
\usepackage{amsthm}
\usepackage{array}
\usepackage{graphicx}
\usepackage[margin=1in]{geometry}
\usepackage{float}
\usepackage{booktabs}
 
\usepackage[config,labelfont=sc,textfont=sl]{caption,subfig}

\usepackage{mathtools}
\usepackage{psfrag}
\usepackage{mathrsfs}
 
\newcommand{\rma}{\mathrm{a}} 
 
\newcommand{\rmc}{\mathrm{c}} 
 
\newcommand{\rme}{\mathrm{e}}

\newcommand{\rmi}{\mathrm{i}}

\newcommand{\rmp}{\mathrm{p}}

\newcommand{\rmx}{\mathrm{x}}  
\newcommand{\rmy}{\mathrm{y}}  
  
\newcommand{\rmB}{\mathrm{B}}

\newcommand{\rmH}{\mathrm{H}} 
 
\newcommand{\rmJ}{\mathrm{J}}

\newcommand{\rmM}{\mathrm{M}}

\newcommand{\rmY}{\mathrm{Y}}

  \newcommand{\bfk}{\mathbf{k}}

      \newcommand{\bfg}{\mathbf{g}}

         \newcommand{\bfP}{\mathbf{P}}            
         \newcommand{\bfQ}{\mathbf{Q}}   
            \newcommand{\bfC}{\mathbf{C}}

\newcommand{\eff}{\mathrm{eff}}

\newcommand{\dbarheps}{\bar{\bar{h}}_\varepsilon}  
\newcommand{\dbarfeps}{\bar{\bar{f}}_\varepsilon}  
 \usepackage{dutchcal}
\newcommand{\emm}{\mathbcal{m}} 

\title{\vspace{-20mm} Two-dimensional Helmholtz resonator arrays. Part II. Matched asymptotic expansions for specially-scaled   resonators}

\author[1]{M. J. A. Smith}
\author[1]{I. D. Abrahams} 

\affil[1]{\small Department of Applied Mathematics and Theoretical Physics, University of Cambridge,
Wilberforce Road, CB3 0WA, UK}

\date{}





\begin{document}
\maketitle
  \vspace{-10mm}
\begin{center}
\small {\it Submitted Manuscript}
\end{center}	
\begin{abstract}
We present   a solution method which combines the method of matched asymptotics with the method of multipole expansions to 
   determine the band   structure  of cylindrical Helmholtz resonators arrays in two dimensions. The resonator geometry is considered in the limit as the wall thickness becomes very large compared with the aperture width (the {\it specially-scaled} limit). In this regime,     the existing treatment in Part I, with updated parameters, is found to return spurious spectral behaviour. We  derive a regularised system which overcomes this issue and also derive compact asymptotic descriptions for the low-frequency dispersion equation in this setting. In the  {\it specially-scaled}  limit, our   asymptotic   dispersion equation  not only recovers the first band surface but   also extends to high, but still  subwavelength, frequencies.  A homogenisation treatment is outlined for describing the effective bulk modulus and effective density tensor of the resonator array for all wall thicknesses. We demonstrate that {\it specially-scaled}  resonators are able to achieve exceptionally low Helmholtz resonant frequencies, and present closed-form expressions for determining these explicitly. We anticipate that the analytical expressions and the   formulation outlined here may prove useful in industrial and other applications.
\end{abstract}

\section{Introduction}  
In Part I of this study, we outlined a matched asymptotic-multipole treatment  for determining the band structure of thin- and moderately thick-walled Helmholtz resonator arrays. However, this formulation    implicitly assumed  that the   wall thickness was not too large as compared with the aperture width; an assumption that prevents us from achieving very low  Helmholtz resonance frequencies, or equivalently, very low first-band gaps. Here we consider an important extension to the  results derived in Part I by examining arrays of {\it specially-scaled} thick-walled resonators (with aspect ratios chosen to achieve a low frequency resonance -- hence the use of the terminology {\it specially-scaled} resonators), see Figure \ref{fig:fundcellthick}. We   also discuss a homogenisation procedure for all wall thickness configurations.

On the topic of homogenisation, we remark that the literature on   two- and three-dimensional  Helmholtz resonator arrays  is extensive, particularly since the resonator can exhibit (1)   singular behaviour with respect to its geometry, as well as (2)  singular behaviour with respect to a constitutive quantity (e.g., high-contrast conductivity). We do not attempt to conduct an exhaustive literature review here but will instead  highlight  key works of interest. To clarify by means of example,  a geometrically singular medium could include an array where the   aperture width of each resonator  contracts much more rapidly than some other geometric parameter,   as wavelengths become long relative to the period of the unit cell (see works in acoustics  \cite{schweizer2017resonance,lamacz2016effective} and in electromagnetism \cite{kohn2008magnetism}).  Interestingly, existing work in acoustics \cite{schweizer2017resonance,lamacz2016effective} suggest that for the {\it specially-scaled} Helmholtz resonator array problem, the effective density is not frequency dependent whereas the effective bulk modulus is indeed frequency dependent and takes negative values for a fixed frequency interval. This very same behaviour is   observed in two-scale asymptotic treatments of  high contrast arrays of cylinders for the Helmholtz equation   \cite{bouchitte2004homogenization,zhikov2005spectrum}. The observation that frequency dependence emerges only in the effective bulk modulus is consistent with the assertion in \cite{haberman2016acoustic} that a dynamic {\it compressibility} is observed near the Helmholtz resonance frequency for Helmholtz resonator arrays, and not a  dynamic {\it density} response.

For those interested in the history of the Helmholtz resonator design, it would seem that the systematic investigation of cavity resonators   emerged in the mid-19th century, after   tone generation was observed in   heated    glass flasks  with long necks  \cite{howe1976helmholtz}. Soon thereafter,   mathematical explanations     were given by Helmholtz \cite{von1860crelle}  and Lord Rayleigh \cite{rayleigh1870on}, the first of these giving rise to the   nomenclature of {\it Helmholtz resonators} that is used   to the present day. The study of two-dimensional resonators  seemingly have their origins in the study of loop antennas  which emerged almost a century later    \cite{schelkunoff1952antennas}, from which the nomenclature of  {\it split-ring resonator}   likewise emerged for the design considered here, as understood by many in the community. For reference, alternative nomenclatures    include  {\it loop-gap resonator} or {\it split-tube resonator} within the electrical engineering literature.

The outline of this paper is as follows. First, we briefly restate the governing equations offered in Part I in the following section. We then determine the leading-order outer and inner solutions within the neck region  in Section \ref{sec:hhrthick}, where asymptotic matching is also conducted. In Section \ref{sec:regmultipole} we construct the regularised system for {\it specially-scaled} resonator arrays before constructing asymptotic dispersion equations in Section \ref{eq:asyregsystem}. In Section \ref{sec:cutoff} we present closed-form representations for the Helmholtz resonance/cutoff frequency. Numerical examples are then considered in Section \ref{sec:numerics}, and finally an extended discussion is given in Section \ref{sec:discus}, which highlights the differences of the model results presented here to those presented in Part I.   

\subsection{Governing wave   equation  }\label{sec:goveq}
\begin{figure}[t]
\centering
\subfloat[Subfigure 6 list of figures text][]{
\includegraphics[width=0.4\textwidth]{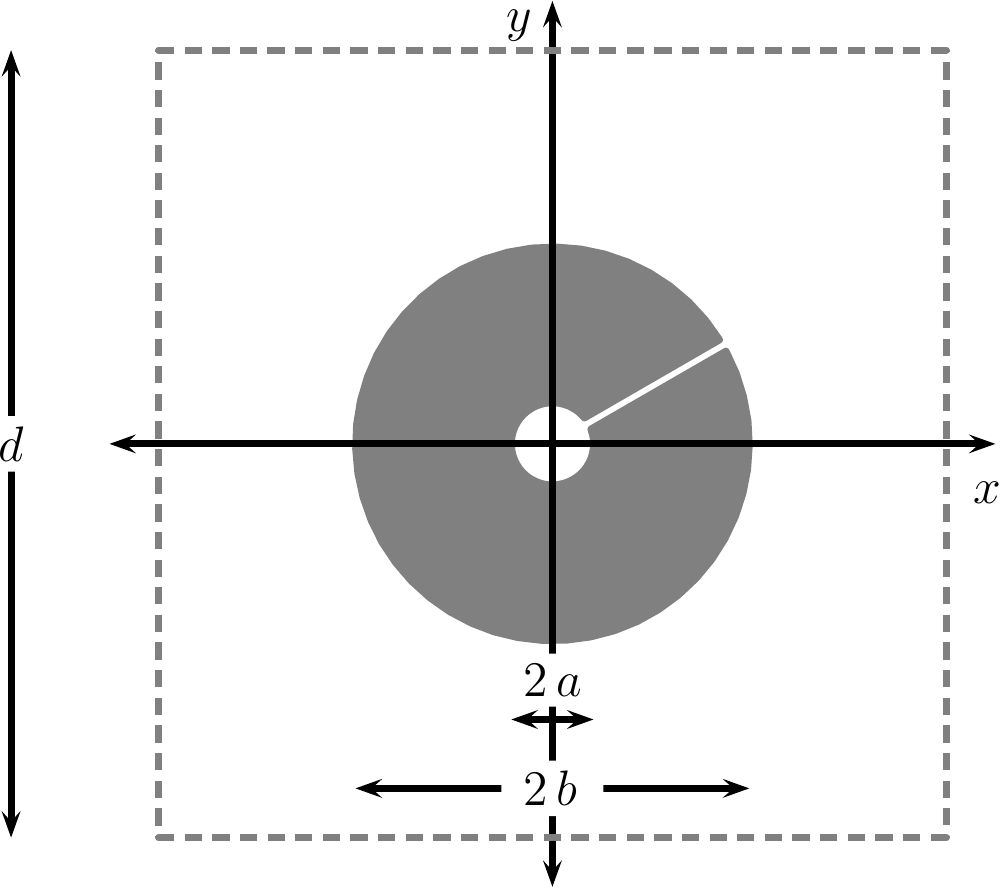}
\label{fig:ucell1}}
\subfloat[Subfigure 1 list of figures text][]{   
\includegraphics[width=0.375\textwidth]{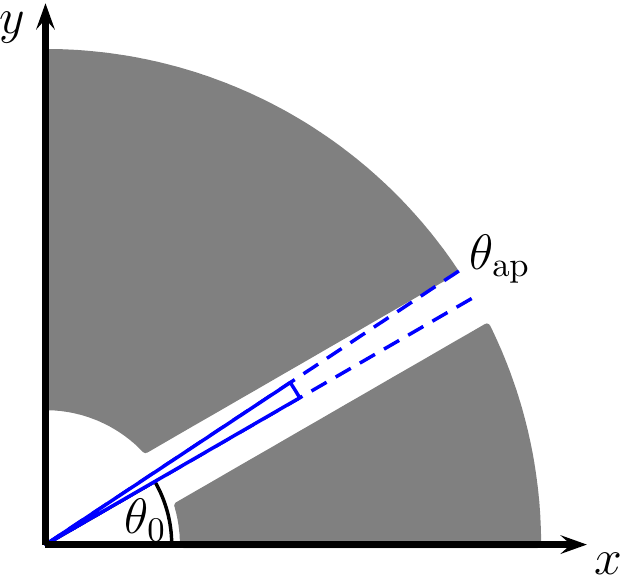} 
\label{fig:ucell2}}

\caption{ \protect\subref{fig:ucell1} Representative fundamental unit cell   in non-dimensional coordinates for a square array of Helmholtz resonators with large aspect ratios (i.e., specially-scaled   Helmholtz resonators), where $a$ and $b$ denote the inner and outer radii, respectively, and $d$ is the periodicity;  \protect\subref{fig:ucell2} Close-up of   neck region showing central aperture angle $\theta_0$ and aperture half-width angle $\theta_\mathrm{ap}$.
 \label{fig:fundcellthick}}
 
\end{figure}

We consider wave propagation in an acoustic  medium satisfying the scalar wave equation  
\begin{equation}
\label{eq:helmnondim}
(\partial_{\overline{x}}^2 + \partial_{\overline{y}}^2)  \, \overline{\phi} +k^2    \overline{\phi} = 0,
\end{equation}
where $\overline{x}$ denotes dimensional Cartesian coordinates,   $k^2 = \rho \omega^2/B$ is the square of the wave number, $B$ is the Bulk modulus,   $\rho$ is the mass density of the medium, and $\bar{\phi}$ is the velocity potential (see Part I for further details). Within this acoustic medium we immerse a two-dimensional square array of  rigid  resonators as shown in Figure \ref{fig:fundcellthick}, spaced a distance $\overline{d}$ apart, satisfying Neumann conditions on the wall edges, and Bloch conditions throughout the cell. Implicitly  we examine time-harmonic solutions of the form $\mathrm{exp}(-\rmi \omega t)$ where $\omega$ is the angular frequency, but this factor is suppressed throughout for ease of exposition.
 
\section{Helmholtz resonators in the specially-scaled limit} \label{sec:hhrthick}
We shall now employ the method of matched asymptotic expansions \cite{crighton1992modern,cotterill2015time} as outlined in Part I, where through nondimensional rescaling we introduced   {\it inner} and {\it outer} regions of the unit cell. However, in contrast to the  treatment of thick-walled resonators in Part I, where we partitioned the unit cell for the outer problem into two    domains, we now partition the unit cell for the outer problem into three domains: the interior, neck, and exterior regions, and consider the solution in each domain. 
The outer solutions for the interior and exterior regions are identical to those presented in Part I, whose results we   restate for   reference below. For the outer domains we use coordinates scaled on the wave number
\begin{equation}
x = k \overline{x}, \quad  \mbox{and}\quad y = k \overline{y},  
\end{equation}
with the inner and outer radii of the resonator, $\overline{a}$ and $\overline{b}$,   scaled as $a = k\overline{a}$ and  $b = k\overline{b}$ respectively, the lattice period $\overline{d}$ is scaled as $d = k\overline{d}$, the nondimensional cylinder thickness is $2\emm=b-a$, and the aperture half-width $\overline{\ell}$ is scaled as $\varepsilon = k\ell$. In all that follows we take the asymptotic limit $\varepsilon \rightarrow 0$. We   now examine  the outer solution in the neck region in   detail. 

\subsection{Outer   solution in the neck region} 
\begin{figure}[t]
\centering
\subfloat[Subfigure 6 list of figures text][]{
\includegraphics[width=0.55\textwidth]{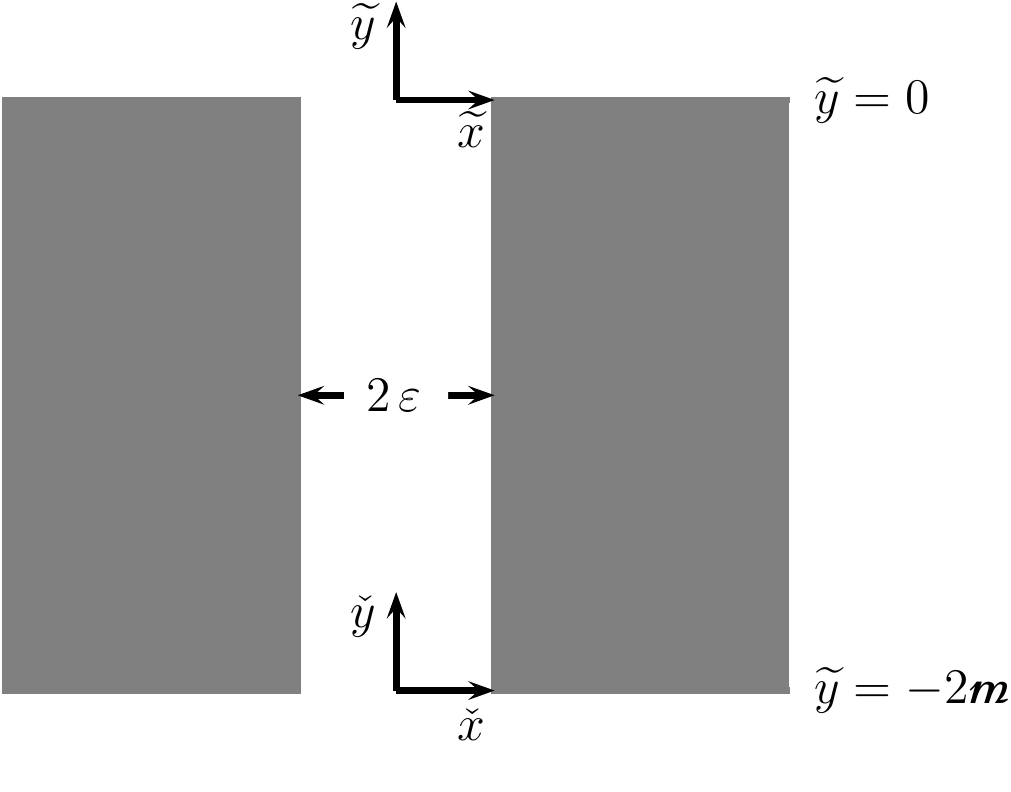}
\label{fig:outerneckthickres1} 
}
\subfloat[Subfigure 1 list of figures text][]{   
\includegraphics[width=0.4\textwidth]{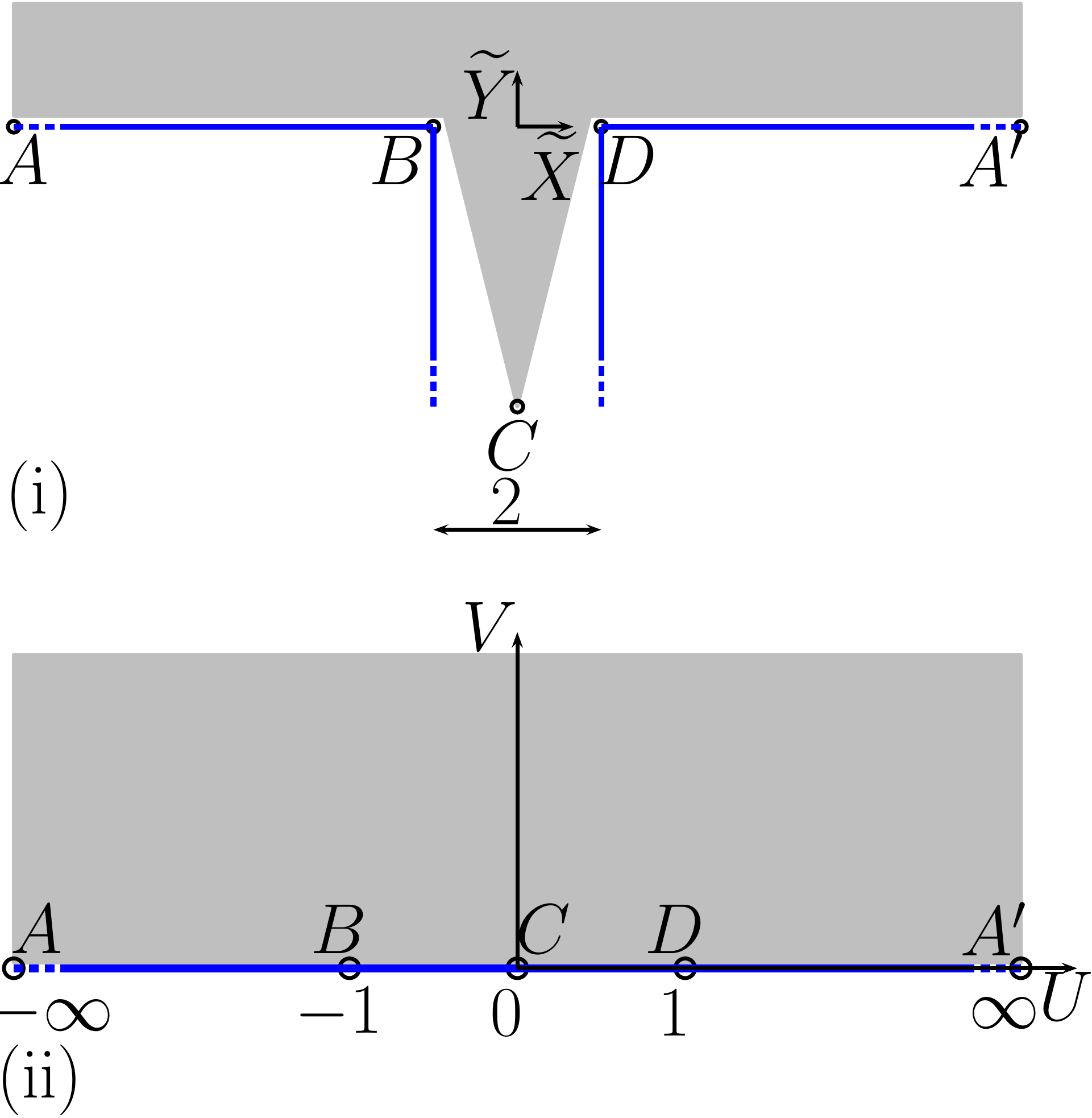} 
\label{fig:outerneckthickres2}}
\caption{ \protect\subref{fig:outerneckthickres1} Outer problem inside the thick-walled resonator neck of width     $2\varepsilon $  and  length     $2\emm$ (not to scale) where $(\widetilde{x},\widetilde{y})$ denotes the centre of the exterior mouth and $(\check{x},\check{y})$ the centre of the interior mouth, \protect\subref{fig:outerneckthickres2}(i) Inner problem geometry at the exterior neck entrance, and \protect\subref{fig:outerneckthickres2}(ii) Inner problem geometry after applying the Schwartz--Christoffel mapping \eqref{eq:scthickscaled}; the capital letters $A,\ldots, D$ and $A^\prime,\ldots, D^\prime$ denote points of correspondence between the two complex planes \label{fig:outerneckthickres}}
\end{figure}

Note that from the unit cell configuration shown in Figure \ref{fig:fundcellthick} we first rotate and translate the lattice as  $(\widetilde{x},\widetilde{y})\mapsto (x\sin\theta_0 - y\cos\theta_0,x\cos\theta_0 + y\sin\theta_0 - b)$, where   $\theta_0$ is the central aperture angle, so that the exterior mouth of the resonator is located at the origin in $(\widetilde{x},\widetilde{y})$ coordinates. In this new coordinate frame  we solve Helmholtz's equation 
\begin{equation}
\label{eq:helmneckoperator}
(\partial_{\widetilde{x}}^2 + \partial_{\widetilde{y}}^2 + 1)\phi_\mathrm{neck} = 0,
\end{equation}
in the neck region of the resonator $S_N = \left\{(\widetilde{x},\widetilde{y}) : (-\varepsilon,\varepsilon) \times(-2\emm ,0)\right\}$ subject to the Neumann boundary conditions
\begin{equation}
\label{eq:neumannouterneck}
\frac{\partial \phi_\mathrm{neck} }{\partial \widetilde{x}}	\bigg|_{\widetilde{x}=\pm\varepsilon} =0, 
\end{equation}
as illustrated in Figure \ref{fig:outerneckthickres}, in the limit as $\varepsilon \rightarrow 0$. Away from the aperture mouths, \eqref{eq:helmneckoperator} and \eqref{eq:neumannouterneck} admit    the general solution
\begin{equation}
\phi_\mathrm{neck}  =  \sum_{n=0}^{\infty} \left\{p_n \rme^{\rmi \lambda_n \widetilde{y}} +q_n \rme^{-\rmi \lambda_n \widetilde{y}}\right\} \cos \left(\frac{n \pi(\widetilde{x}+\varepsilon)}{2\varepsilon} \right), \qquad
\end{equation}
where
\begin{equation}
\lambda_n = 
\begin{cases}
\sqrt{1 - \left( \dfrac{n \pi}{2\varepsilon}\right)^2}, & \dfrac{n \pi}{2\varepsilon}<1,\\
\rmi\sqrt{\left( \dfrac{n \pi}{2\varepsilon}\right)^2 - 1}, & \dfrac{n \pi}{2\varepsilon}>1.
\end{cases}
\end{equation}
 In the closing aperture limit, the dominant contribution comes from the $n=0$ term and so the solution takes the form
 \begin{equation}
 \label{eq:outerneckphi}
\lim_{\varepsilon\rightarrow 0}\phi_\mathrm{neck} \sim p_0 \rme^{\rmi   \widetilde{y}} +q_0 \rme^{-\rmi   \widetilde{y}}.
\end{equation}
Accordingly the outer solution asymptotics near the entrance and exit to the neck are given by
\begin{subequations}
\label{eq:neckbothexp}
\begin{align}
\label{eq:neckouterasy1}
\lim_{\widetilde{x}\rightarrow 0}\lim_{\widetilde{y}\rightarrow 0} \phi_\mathrm{neck} &\sim (p_0 + q_0) + \rmi \widetilde{y} (p_0 - q_0), \\
\label{eq:neckouterasy2}
\lim_{\check{x}\rightarrow 0}\lim_{\check{y}\rightarrow 0} \phi_\mathrm{neck} &\sim \left(p_0 \rme^{-2\rmi \emm} + q_0 \rme^{2\rmi \emm} \right)+ \rmi \check{y} (p_0 \rme^{-2\rmi \emm} - q_0 \rme^{2\rmi \emm}),
\end{align}
\label{eq:outerasyboth}
\end{subequations}
 respectively, where    we express \eqref{eq:neckouterasy2} in terms of the shifted origin $(\check{x},\check{y}) = (\widetilde{x},\widetilde{y}+2\emm)$. We note that for the inner solutions that follow, we use the  same rotated and translated frame $(\widetilde{x},\widetilde{y})$, and so we do not need to express the representations \eqref{eq:outerasyboth} above in terms of $(x,y)$.

\subsection{Outer solution in the interior and exterior regions}
Next, we restate results for the outer solution in the interior and exterior domains (see Eq.~(6.11) of Part I) as we approach the mouths in the form
\begin{subequations}
\label{eq:outerasympsbothdirthick}
\begin{align} \nonumber
\lim_{\theta \rightarrow \theta_0} \lim_{r\rightarrow b} \phi_\mathrm{ext} &\sim
 \dfrac{2\rmi A}{\pi} \left[ \gamma_\rme - \dfrac{\rmi \pi}{2} + \log\left(\dfrac{\widetilde{r}}{2}\right) \right]  + \sum\limits_{n=-\infty}^{\infty}b_n    \rmY_n(b)     \rme^{\rmi n \theta_0}
  \\  
&\hspace{30mm}-\sum\limits_{n=-\infty}^{\infty} \left\{     \dfrac{A Q_n}{2  }    + b_n \rmY_n^\prime(b)    \rme^{ \rmi n    \theta_0 } \right\}   \dfrac{\rmJ_n(b)}{\rmJ_n^\prime(b)}, \\
 \lim_{\theta \rightarrow \theta_0} \lim_{r\rightarrow a} \phi_\mathrm{int}  &\sim
\dfrac{2\rmi B}{\pi} \left[ \gamma_\rme - \dfrac{\rmi \pi}{2} + \log\left(\dfrac{\check{r}}{2}\right) \right] - \dfrac{B}{2}\sum\limits_{n=-\infty}^{\infty} \dfrac{\check{Q}_n}{\rmJ_n^\prime(a)} \rmJ_n(a),  	 
\end{align}
\end{subequations} 
where  
   $Q_n  =  	\rmJ_n(b)\rmH_n^{(1)\prime}(b) + \rmJ_n^\prime(b) \rmH_n^{(1)}(b)$ and 
 $ \check{Q}_n =  	\rmJ_n(a)\rmH_n^{(1)\prime}(a) + \rmJ_n^\prime(a) \rmH_n^{(1)}(a)$,
along with $\widetilde{r} = \sqrt{(x-b\cos\theta_0)^2+(y-b\sin\theta_0)^2}$, $\check{r} = \sqrt{(x-a\cos\theta_0)^2+(y-a\sin\theta_0)^2}$, and $a=b-2\emm$. In the above, $\rmJ_n(z)$ and $\rmY_n(z)$ are Bessel functions of the first and second kind, respectively, and $\rmH_n^{(1)}(z)$ are Hankel functions of the first kind. Having obtained   asymptotic forms for the outer solution in the interior, neck, and exterior regions, we now consider the task of determining inner solutions at the junctions to the resonator neck.

\subsection{Inner solutions and asymptotic matching procedure}
 As before, we   first rotate and translate the lattice  so that the exterior mouth of the resonator is located at the origin in $(\widetilde{x},\widetilde{y})$ coordinates. We now   introduce the inner   scaling 
$
\widetilde{X} = \widetilde{x}/\varepsilon$ and  $\widetilde{Y} = \widetilde{y}/\varepsilon$ 
along with   a regular expansion for $\phi$ as in Part I; substituting these into the   system we obtain the leading-order inner boundary value problem
\begin{subequations}
\begin{align}
( \partial_{\widetilde{X}}^2 + \partial_{\widetilde{Y}}^2) \Phi = 0, \quad \mbox{ for } \widetilde{X}  \in   \widetilde{S}_\rmM, \\
\partial_{\widetilde{N}} \Phi  = 0, \quad \mbox{ for } \widetilde{X}  \in \partial \widetilde{S}_\rmM,
\end{align}
\end{subequations}
where   $\partial_{\widetilde{N}}$ denotes the normal derivative, $\widetilde{S}_\rmM =  \{(\widetilde{X} ,\widetilde{Y} ): -\infty<\widetilde{Y}<0, |\widetilde{X}|\leq 1  \} \cup   \{(\widetilde{X} ,\widetilde{Y} ): \widetilde{Y}>0  \}$ is the exterior mouth domain shown in Figure \ref{fig:outerneckthickres}, and for clarity we   omit the subscript for $\Phi_0$, which denotes the leading term in the inner expansion as $\varepsilon\rightarrow 0$. To obtain a solution we map the exterior mouth region  $\widetilde{S}_\rmM$ to the upper-half plane via the Schwartz--Christoffel mapping
\begin{equation}
\label{eq:scthickscaled}
\widetilde{Z}(W) = \frac{2}{\pi} \left[ (W^2-1)^{1/2} - \rmi \log\left(  (W^2-1)^{1/2} + \rmi\right) + \rmi \log W \right],
\end{equation}
where $\widetilde{Z} = \widetilde{X} + \rmi \widetilde{Y}$ and  $W = U+\rmi V$, 
which   exhibits the asymptotic behaviours
\begin{subequations}
\begin{align}
\lim_{W\rightarrow 0} \widetilde{Z} (W)  \sim 1 + \frac{2\rmi}{\pi} (1-\log 2) + \frac{2\rmi}{\pi} \log W, \quad \mbox{ and } \quad
\lim_{W\rightarrow \infty}\widetilde{Z} (W)  \sim \frac{2 W}{\pi}.
\end{align}
\label{eq:Wasy}
\end{subequations}
The appropriate solution to Laplace's equation in the upper-half plane satisfying Neumann conditions along $V=0$ is   given by
\begin{equation}
\Phi = C_1 \mathrm{Re}(\log W) + C_2,
\end{equation}
where $C_1$ and $C_2$ are as yet unknown and from \eqref{eq:Wasy} it follows that
\begin{subequations}
\label{eq:asyinnerbothmouths}
\begin{align}
\lim_{\substack{\widetilde{Z} \rightarrow \infty \\ (UHP)}} \Phi  = C_1 \log \left(\frac{\pi \widetilde{R} }{2} \right) + C_2, \quad \mbox{ and } \quad
\lim_{\substack{\widetilde{Y} \rightarrow -\infty \\ |\widetilde{X} | < 1}} \Phi  = C_1\left[ \frac{\pi \widetilde{Y} }{2} - 1 + \log 2\right]+ C_2.
\end{align}
\label{eq:Phiasy}
For the   interior mouth region, the inner solution $\Psi$ is obtained by a treatment   analogous to that outlined above, but   now expressed in terms of the inner transformation
$
\check{X} =  \check{x}/\varepsilon
$
and
$
\check{Y} =  \check{y}/\varepsilon  = (\widetilde{y} + 2\emm)/\varepsilon
$.
Accordingly, in this region, from \eqref{eq:scthickscaled}, we expect  the asymptotic behaviour
\begin{align}
\lim_{\substack{\check{Z}\rightarrow \infty \\ (LHP)}} \Psi  = C_3 \log \left(\frac{\pi \check{R}}{2} \right) + C_4, \quad \mbox{ and } \quad
\lim_{\substack{\check{Y}\rightarrow \infty \\ |\check{X}| < 1}} \Psi  = C_3\left[ -\frac{\pi \check{Y}}{2} - 1 + \log 2\right]+ C_4,
\end{align}
\label{eq:Phiasy2}
\end{subequations}
where $\check{Z} = \check{X} + \rmi \check{Y} = \check{R} \, \mathrm{exp}(\rmi \check{\Theta} )$ and both $C_3$ and $C_4$ are as yet unknown. Subsequently, the matching procedure \cite{crighton1992modern}, at leading order, gives rise to the relations
\begin{subequations}
\label{eq:matchingconds}
\begin{align}
\label{eq:matching1}
\lim_{\substack{ \widetilde{R} \rightarrow \infty \\ (UHP)}} \Phi \bigg|_{\widetilde{R} = \widetilde{r}/\varepsilon} = \lim_{\widetilde{r}\rightarrow 0}\lim_{\theta \rightarrow \theta_0}\phi_\mathrm{ext}, 
\quad&\mbox{and}\quad 
\lim_{\widetilde{Y}\rightarrow -\infty} \Phi \bigg|_{\widetilde{X} = \widetilde{x}/\varepsilon, \widetilde{Y} = \widetilde{y}/\varepsilon  } \qquad  = \lim_{\widetilde{x}\rightarrow 0} \lim_{  \widetilde{y}\rightarrow 0} \phi_\mathrm{neck},
\\
\label{eq:matching4}
\lim_{\substack{\check{R}\rightarrow \infty \\ (LHP)}} \Psi \bigg|_{\check{R} = \check{r}/\varepsilon} = \lim_{\check{r}\rightarrow 0}\lim_{\theta \rightarrow \theta_0}\phi_\mathrm{int},
\quad&\mbox{and}\quad 
\lim_{\check{Y}\rightarrow  \infty}  \Psi \bigg|_{\check{X}= \check{x}/\varepsilon, \check{Y}  = \check{y}/\varepsilon  }  = \lim_{\check{x}\rightarrow 0} \lim_{  \check{y}\rightarrow 0} \phi_\mathrm{neck}.
\end{align}
\end{subequations}
 Thus,   matching polynomial orders between \eqref{eq:neckbothexp} and \eqref{eq:asyinnerbothmouths}, in addition to logarithmic and non-logarithmic terms between \eqref{eq:outerasympsbothdirthick} and   \eqref{eq:asyinnerbothmouths}, allows us to determine, after significant algebra, all coefficients $C_j$, as well as the monopole amplitudes $B = 2\rmi A\, (2\rmi \tau_1/\pi + \tau_2 \tau_5)^{-1}/\pi$  and 
\begin{equation}
\label{eq:crucialA}
 A     =    \frac{2}{\pi b  \dbarheps}\sum\limits_{n=-\infty}^{\infty} g_n,
\end{equation}
where $g_n  =  b_n   \rme^{\rmi n \theta_0}  / \rmJ_n^\prime(b)$, and
\begin{equation}
\label{eq:hepsbigasp}
\dbarheps  =  \dfrac{2\rmi  }{\pi} \left[ \gamma_\rme - \dfrac{\rmi \pi}{2}   -  \log\left(\frac{\pi}{\varepsilon}\right)-\left( \frac{2\rmi}{\pi} \tau_3  + \tau_4 \tau_5    \right)\left( \frac{2\rmi}{\pi} \tau_1 + \tau_2 \tau_5 \right)^{-1}\right] 
 - \frac{1}{2}  \sum\limits_{n=-\infty}^{\infty}    \dfrac{Q_n \rmJ_n(b)}{  \rmJ_n^\prime(b) }     ,
\end{equation}
along with
\begin{subequations}
\label{eq:listotaus}
\begin{align}
\tau_1 &=  \frac{2\varepsilon}{\pi} (1-\log 2) \sin(2\emm) - \cos(2\emm), \qquad 
\tau_4  =  -\frac{2\varepsilon}{\pi} (1 - \log 2) \sin(2\emm)+ \cos(2\emm), 
\\
\tau_2 &= -\frac{2\varepsilon}{\pi}\sin(2\emm), \hspace{20mm}
\tau_5  = \dfrac{2\rmi  }{\pi} \left[ \gamma_\rme - \dfrac{\rmi \pi}{2}  -\log\left(\frac{\pi}{ \varepsilon}\right) \right] - \dfrac{1}{2}\sum\limits_{n=-\infty}^{\infty} \dfrac{\check{Q}_n}{\rmJ_n^\prime(a)} \rmJ_n(a)  ,
\\
\tau_3 &=  \left[ \frac{2\varepsilon}{\pi} (1-\log 2)^2   - \frac{\pi}{2\varepsilon}\right]\sin(2\emm) - 2 (1-\log 2)  \cos(2\emm).  
\end{align}
\end{subequations}
Hence we obtain the same multipole eigensystem as in Eq.~(4.16) of  Part I, which we express as
  \begin{multline}
\label{eq:dispeqsystemgnptii}
\dfrac{\rmi A}{2} \left(    \frac{\rmJ_n^\prime(b)\rmY_n(b) + \rmY_n^\prime(b)   \rmJ_n(b)}{\rmJ_n^\prime(b) \rmY_n^\prime(b)}  +     2 \sum_{m=-\infty}^{\infty} (-1)^{n+m}S_{m-n}^\rmY(\bfk_\rmB) \frac{\rmJ_m(b)}{\rmY_n^\prime(b)} \rme^{-\rmi (m-n)  \theta_0}  \right) \\
 + g_n +      \sum_{m=-\infty}^{\infty} (-1)^{m+n}S_{m-n}^\rmY(\bfk_\rmB) \frac{\rmJ_m^\prime(b) }{\rmY_n^\prime(b)} \rme^{-\rmi(m-n)\theta_0}	g_m  = 0  ,
\end{multline} 
 but with the replacement $h_\varepsilon\mapsto \dbarheps$ as written in \eqref{eq:crucialA}  above. Thus, it would appear that we have an eigensystem for examining {\it specially-scaled}   resonator arrays, however, it turns out that the above   formulation  exhibits pathological behaviour at frequencies near $\dbarheps=0$ which occurs at much lower frequencies than for the wall thickness values discussed in Part I. We now present a regularisation procedure for resolving this issue.

\section{Regularised multipole system formulation} \label{sec:regmultipole}
As indicated in the previous section, although the   system  in Part I with the replacement $h_\varepsilon\mapsto \dbarheps$ is indeed formally correct,  it can return spurious spectral behaviours upon truncating for numerical evaluation (i.e., we may observe  incorrect folded bands, and flat band surfaces, that are not part of the genuine spectrum   in a neighbourhood around frequencies corresponding to $\dbarheps=0$).  We easily resolve this issue by   summing \eqref{eq:dispeqsystemgnptii} over all $n$ to obtain a   more numerically stable yet formally equivalent representation   for $A$ (c.f., to the form \eqref{eq:crucialA} above) as
\begin{equation}
\label{eq:Anewparttwo}
A = 2\rmi \left( E  + 2 \left[ \sum_{v=-\infty}^{\infty} J_v(b) F_v \right] - \rmi \pi b \dbarheps
\right)^{-1} \sum_{m=-\infty}^{\infty}  \rmJ_m^\prime(b) F_m g_m,
\end{equation} 
where
 \begin{equation}
 \label{eq:EandFforms}
E =   \sum_{u=-\infty}^{\infty} E_u,
\quad \mbox{and}\quad
F_m = \sum_{p=-\infty}^{\infty} \dfrac{(-1)^{p-m} S_{m-p}^\rmY(\bfk_\rmB) \rme^{\rmi (p-m) \theta_0}}{\rmY_{p}^\prime(b)}
,	
\end{equation} 
  with $E_u = \left[\rmJ_u^\prime(b)\rmY_u(b) + \rmY_u^\prime(b)   \rmJ_u(b)\right]/\left[\rmJ_u^\prime(b) \rmY_u^\prime(b)\right]$. Substituting this new representation  for $A$   \eqref{eq:Anewparttwo}   into \eqref{eq:dispeqsystemgnptii}   admits the regularised system
  \begin{equation}
\label{eq:dispeqsystemgnptfinal}
  g_n +      \sum_{m=-\infty}^{\infty} (-1)^{m+n}S_{m-n}^\rmY(\bfk_\rmB) \frac{\rmJ_m^\prime(b) }{\rmY_n^\prime(b)} \rme^{-\rmi(m-n)\theta_0}	g_m  -\dfrac{\chi_n}{H_\varepsilon} 
      \sum_{m=-\infty}^{\infty}  \rmJ_m^\prime(b) F_m g_m   = 0  ,
\end{equation} 
for all $n$, where
\begin{subequations}
\begin{equation}
\label{eq:chinregform}
 \chi_n =  E_n +     2 \sum_{p=-\infty}^{\infty} (-1)^{n+p}S_{p-n}^\rmY(\bfk_\rmB) \frac{\rmJ_p(b)}{\rmY_n^\prime(b)} \rme^{-\rmi (p-n)  \theta_0},
\end{equation} 
and  
\begin{equation}
\label{eq:Hepsregform}
H_\varepsilon =  E  + 2 \left[ \sum_{v=-\infty}^{\infty} J_v(b) F_v \right] - \rmi \pi b \dbarheps.
\end{equation}
\end{subequations}
On examining the   system   \eqref{eq:dispeqsystemgnptfinal} numerically, we find that spurious      effects are removed and the genuine spectrum is   observed, see for example Figure \ref{fig:foldedboth} which is discussed in further detail below. In order for the regularisation to be effective,  we remark that the known expressions $E$, $\chi_n$, $F_m$, and $H_\varepsilon$ must be suitably converged with appropriately chosen truncation numbers for the sums, which we represent by $L_E$, $L_{\chi_n}$, $L_{F_m}$, and $L_{H_\varepsilon}$.

\section{Asymptotic   representations for the dispersion equation } \label{eq:asyregsystem}
In this section, we construct an   asymptotic representation  of the dispersion equation  for the  regularised system \eqref{eq:dispeqsystemgnptfinal} at low frequencies. We begin by observing that the lattice sums $S_m^\rmY$, which   feature  in   \eqref{eq:dispeqsystemgnptfinal}    as well as in the   functions $F_m$   \eqref{eq:EandFforms}, $\chi_n$   \eqref{eq:chinregform}, and $H_\varepsilon$   \eqref{eq:Hepsregform}, can be expressed as  

\begin{equation}
\label{eq:SlYexpgeneral}
S_m^\rmY \approx \sum_{r =-\Omega_m}^{\infty} \beta_r^{(m)} b^r,
\end{equation}
where  
$
\Omega_m =  2\left[  \mathrm{Floor}\left\{(m-3)/4 \right\} +\mathrm{Floor}\left\{(m-4)/4 \right\} +2\right]
$
for $m\geq 4$,
  denotes the first non-zero order of the lattice sum (i.e.,  $\Omega_4 = 4$ and $\Omega_7=6$), along with $\Omega_0 = \Omega_1 = \Omega_2 = \Omega_3 = 2$ \cite[A004524]{oeisA004524}. Closed-form expressions for the first few lattice sums, i.e., the values $\beta_r^{(m)}$ for small $m$, are derived and presented in Appendix \ref{sec:asymptoticsSmY} for reference. Note that these terms may contain logarithmic behaviour in the lattice spacing $d$ and therefore in $b$ if the area fraction $f=\pi b^2/d^2$ is fixed. With the lattice sum  expansions \eqref{eq:SlYexpgeneral},   we examine the system  \eqref{eq:dispeqsystemgnptfinal}   in the limit of vanishing $b$ and to within a dipole truncation in $n$, which admits the matrix form   
\begin{equation}
\label{eq:sysmatrixrep}
(H_\varepsilon \bfC - \bfP -\bfQ)\bfg \approx \boldsymbol{0},
\end{equation}
where $\bfC$ is the matrix representation of the system for an array of Neumann cylinders (obtained by setting $\chi_n=0$   in  \eqref{eq:dispeqsystemgnptfinal}, see \cite[Eq.~(3.120)]{movchan2002asymptotic}), $\bfP$ and $\bfQ$ are    perturbation matrices due to the (specially-scaled) aperture, and $\bfg$ is the vector of $g_m$ coefficients. These matrices take the forms
\begin{subequations}
\begin{equation}
\bfC \approx 
\left[
\begin{array}{c|c|c} \phantom{\bigg|^1}
 1 +\dfrac{\pi}{4}   \beta _{-2}^{(0)} 
 & 
  \dfrac{\pi  b }{4}  \rme^{\rmi \theta_0} \left(\beta _{-2}^{(1)}\right)^\ast 
  &
  -\dfrac{\pi}{4}   \rme^{2 \rmi \theta_0} \left(\beta _{-2}^{(2)}\right)^* 
   \\ \hline \phantom{\bigg|^1}
 -\dfrac{\pi  }{4 b} \rme^{-\rmi \theta_0} \beta _{-2}^{(1)}
    &
  1-\dfrac{\pi}{4}   \beta _{-2}^{(0)}
    &
  \dfrac{\pi  }{4 b} \rme^{\rmi \theta_0} \left(\beta _{-2}^{(1)}\right)^* 
     \\ \hline
    -\dfrac{\pi}{4}   \rme^{-2 \rmi \theta_0}  \beta _{-2}^{(2)} 
     &
   -\dfrac{\pi  b }{4}  \rme^{-\rmi \theta_0} \beta _{-2}^{(1)} 
   &      
1+\dfrac{\pi}{4}   \beta _{-2}^{(0)} 
\end{array}
\right],
\end{equation}
\begin{equation}
\bfP \approx 
\left[
\begin{array}{c|c|c}   \phantom{\bigg|^1}
\dfrac{\pi^2 \left| \beta _{-2}^{(1)}\right|^2}{4b} 
 & 
\dfrac{\pi^2 \rme^{\rmi \theta_0} (\beta _{-2}^{(1)})^\ast \beta _{-2}^{(0)}}{4}
  &
-\dfrac{\pi^2\rme^{2\rmi\theta_0}((\beta _{-2}^{(1)})^\ast)^2}{4b} 
   \\ \hline \phantom{\bigg|^1}
-\dfrac{\pi  \left(\pi  \beta _{-2}^{(0)}-2\right) \rme^{-\rmi \theta_0} \beta _{-2}^{(1)} }{4 b^{2}} 
    &
 -\dfrac{\pi  \beta _{-2}^{(0)} \left(\pi  \beta _{-2}^{(0)}-2\right)}{4 b}
    &
 \dfrac{\pi  \left(\pi  \beta _{-2}^{(0)}-2\right)  \rme^{\rmi \theta_0} (\beta _{-2}^{(1)})^\ast}{4 b^{2}} 
   \\ \hline \phantom{\bigg|^1}
-\dfrac{ \pi^2\rme^{-2\rmi\theta_0}( \beta _{-2}^{(1)})^2}{4b} 
     &
-\dfrac{ \pi^2 \rme^{-\rmi\theta_0} \beta _{-2}^{(0)} \beta _{-2}^{(1)}}{4} 
   &      
\dfrac{\pi^2 \left| \beta _{-2}^{(1)}\right|^2}{4b}  
\end{array}
\right],
\end{equation}
and
\begin{equation}
\bfQ \approx 
\left[
\begin{array}{c|c|c} \phantom{\bigg|^1}
 \dfrac{ \pi^2 \rme^{-\rmi \theta_0}\beta _{-2}^{(1)} }{2}\eta^\ast +\rme^{\rmi\theta_0} \pi^2 (\beta _{-2}^{(1)})^\ast\eta
 & 
    \dfrac{ \pi^2 b}{2}\beta _{-2}^{(0)} \eta^\ast  
  &
   \dfrac{\rme^{\rmi\theta_0} \pi^2 (\beta _{-2}^{(1)})^\ast}{2} \eta^\ast
   \\ \hline \phantom{\bigg|^1}
    \dfrac{\pi \eta}{  b}   \left(2 - \pi  \beta _{-2}^{(0)}   \right)  
    &
    0
    &
     \dfrac{\pi \eta^\ast}{  b}   \left(2 - \pi  \beta _{-2}^{(0)}   \right)  
     \\ \hline
     -\dfrac{\pi^2  \rme^{-\rmi\theta_0}\beta _{-2}^{(1)}}{2} \eta 
     &
     \dfrac{ \pi^2 b     }{2} \beta _{-2}^{(0)}\eta 
   &      
-\dfrac{ \pi^2\rme^{ \rmi \theta_0} (\beta _{-2}^{(1)})^\ast }{2} \eta - \rme^{-\rmi\theta_0} \pi^2   \beta _{-2}^{(1)}  \eta^\ast 
\end{array}
\right],
\end{equation}
\end{subequations}  
where we remark that  next order terms     are taken in most matrix entries above for our analysis in the following section,  but are omitted here  for compactness,   and
\begin{equation}
\label{eq:etaexpl}
\eta =  -\frac{\beta _{-2}^{(0)}}{4} + \frac{1}{4} \rme^{-2\rmi\theta_0}  \beta _{-2}^{(2)}  + \sum_{q=1}^{\infty} \left\{ \frac{ \beta_{-2q-2}^{(2q+2)}  \rme^{- (2q+2) \rmi \theta_0}}{4^{q+1}(2q+1)!} \right\},
\end{equation}
 with $\ast$ representing the complex conjugate operation. In addition, we have
 \begin{equation}
H_\varepsilon \approx
\frac{-2 +4 \dbarfeps + \pi \beta _{-2}^{(0)} }{b} - \frac{\pi}{2} \left[ \rme^{\rmi \theta_0} (\beta _{-2}^{(1)})^\ast - \beta _{-2}^{(1)} \rme^{-\rmi\theta_0}\right]  + O(b),
\end{equation}
where we take an analogous scaling to that found in Part I 
\begin{subequations}
\begin{equation}
\label{eq:fepsbigasy}
\dbarfeps = \frac{\pi b^2 }{4\rmi} \dbarheps,
\end{equation} 
as well as  a dipole truncation in $\dbarheps$  \eqref{eq:hepsbigasp} to obtain the   asymptotic form
\begin{multline}
\label{eq:hepsasybigasp}
\lim_{b\rightarrow 0}\lim_{a\rightarrow 0}  \lim_{\emm\rightarrow 0} \dbarheps \sim  \dfrac{2\rmi  }{\pi} \left\{  \frac{1}{b^2} - \frac{1}{8}  - \log\left(\frac{ \pi b}{2\varepsilon}\right)   \right. \\ \left.
 + \left[   \frac{1}{  a^2 } -    \frac{ \emm \pi  }{\varepsilon} - \frac{17}{8} -  \log\left(\frac{\pi a}{8\varepsilon}\right)    \right]\left[ 1 + \frac{4\varepsilon \emm}{\pi} \left(  \frac{1}{  a^2 } -\frac{1}{8} - \log\left(\frac{\pi a}{2\varepsilon}\right)     \right) \right]^{-1}\right\}.
\end{multline}
\end{subequations}
With  the above expansions, we are now able to construct  closed-form representations for the dispersion equation of {\it specially-scaled} resonators, however before proceeding to this task, we comment   that    certain $\beta_{-2n}^{(2n)}$ terms in \eqref{eq:etaexpl} are vanishing for square lattices, such as $\beta_{-6}^{(6)}$, $\beta_{-10}^{(10)}$, and $\beta_{-14}^{(14)}$ and that many terms are readily extracted from the  explicit forms   in Appendix \ref{sec:asymptoticsSmY}.

 \subsection{Dispersion equation forms along selected symmetry directions} \label{sec:disprelsstwhelmres}
 It is instructive  to consider cases where the resonator geometry has natural symmetry. Thus, with the asymptotic system   \eqref{eq:sysmatrixrep} in mind, we now  take   $\theta_0 = 0$ and consider Bloch coordinates located on the high symmetry planes of the Brillouin zone, where $(k_\rmB,\theta_\rmB)$ denotes the polar form of the dimensionless Bloch vector $\bfk_\rmB$. 
 
 For reference, an outline of the symmetry planes for the fundamental cell  in reciprocal space (Brillouin zone) is presented in Figure 1 of Part I, showing the paths between the high symmetry  Bloch vector  coordinates $\Gamma = (0,0)$, $X=(\pi/d,0)$, $M=(\pi/d,\pi/d)$, and $Y=(0,\pi/d)$.  In the first instance, we evaluate the determinant of the system \eqref{eq:sysmatrixrep} along the $\Gamma X$ (i.e., $\theta_\rmB = 0$) direction, which to leading order, returns  the isotropic  result from Part I (Eq.~(4.22)) with the replacement $f_\varepsilon \mapsto \dbarfeps$ in the form
 \begin{equation}
\label{eq:dispeqdipoleleadinginpt2}
k_\rmB^2 = \frac{1+f}{1-f} \left(1 -  \frac{   2 f (1 - \dbarfeps )}{1 - 2\dbarfeps} \right),
\end{equation} 
where   $f = \pi b^2/d^2$ denotes the area ratio and $\dbarfeps$ is given in \eqref{eq:fepsbigasy}.    In contrast, along the $\Gamma Y$ direction ($\theta_\rmB = \pi/2$) we observe a much more complicated form, which emerges from the   symmetries of the system matrices
 \begin{equation*}
\bfC = 
\left[\begin{array}{rrr}
\bullet & \bowtie &\boxtimes \\
\diamond & \circledcirc & -\diamond \\
\boxtimes & -\bowtie &\bullet
\end{array} \right], \quad
\bfP = 
\left[\begin{array}{rrr}
 \bullet & \bowtie &-\bullet \\
\diamond & \circledcirc & -\diamond \\
-\bullet & -\bowtie &\bullet
\end{array} \right], \quad
\bfQ = 
\left[\begin{array}{ccc}
 \bullet & \bowtie &-\boxtimes \\
\diamond & 0 &  \diamond \\
\boxtimes & \bowtie &-\bullet
\end{array} \right], 
\end{equation*}
where symbols are used to represent the symmetry of each matrix {\it independently} (and does not imply equivalence of values between matrices). Subsequently, after a significant amount of algebra, we obtain the dispersion equation 
\begin{multline}
\label{eq:disprelscalGY}
\left(  \sqrt{\frac{ a^2 \left(2 f+(1-2 \dbarfeps) k_\rmB^2\right) +b^2 \left(k_\rmB^2+1\right) -\Lambda_{\Gamma \rmY}}{2 a^2 (1-2 \dbarfeps)+b^2}}-1\right) \cdot
\\
\left(  \sqrt{\frac{a^2 \left(f^2+f (2 \dbarfeps-1) k_\rmB^2+f+(1-2 \dbarfeps) k_\rmB^2\right)+b^2 \left(-f^2-f k_\rmB^2+k_\rmB^2+1\right)+\xi_{\Gamma \rmY}}{2\left(1-f^2\right) \left(a^2 (1-2 \dbarfeps)+b^2\right)}}-1\right)=0,
\end{multline}
\begin{subequations}
where 
\begin{multline}
\Lambda_{\Gamma \rmY}^2 =  a^4  (2 f+(1-2 \dbarfeps) k_\rmB^2 )^2+2 a^2 b^2  [2 f  (k_\rmB^2+1 )+(1-2 \dbarfeps) \left(k_\rmB^2-1\right) k_\rmB^2 ]+b^4  (k_\rmB^2-1 )^2,
\end{multline} 
and
\begin{multline}
\xi_{\Gamma \rmY}^2= \left[a^2 \left(f^2+f (2 \dbarfeps-1) k_\rmB^2+f+(1-2 \dbarfeps) k_\rmB^2\right)+b^2 \left(-f^2-f k_\rmB^2+k_\rmB^2+1\right)\right]^2 \\
-4 b^2 (f-1) \left(f^2-1\right) k_\rmB^2 \left(a^2 (1-2 \dbarfeps)+b^2\right).
\end{multline}
\end{subequations}
The key point here is that although the dispersion equation along $\Gamma \rmY$ in  \eqref{eq:disprelscalGY} above is  indeed accurate, it is however    intractable and does not provide  any useful  insight, despite its correspondence to  the     high symmetry settings of $\theta_0=0$ and $\theta_\rmB=\pi/2$. Attempts   in reducing  \eqref{eq:disprelscalGY}, for example, by taking a Taylor series expansion in small $k_\rmB$, provide equally unhelpful forms. Despite its length, the expression \eqref{eq:disprelscalGY} does however demonstrate that incorporating anisotropy to the   dispersion equation for Helmholtz resonator arrays {\it in the specially-scaled limit} is a much more formidable task than for the (moderately)  thin-walled case, as    shown in Part I.  

The unexpected complexity of the dispersion equation \eqref{eq:disprelscalGY} can be understood by considering the geometry of the resonator: in the {\it specially-scaled} limit, the neck of each resonator is so   thin and long that the resonator is almost invariant under all rotation and reflection operations for the square lattice. Subsequently the metamaterial may be considered {\it almost-isotropic}, even as we approach the resonance frequency, and so next-order asymptotic corrections take on much more complicated forms. Numerical investigations confirm much smaller anisotropy in this scaled setting, to the point where the medium may be essentially regarded   as isotropic for  practical applications.  Accordingly,   the low-frequency dispersion equation for these specially-scaled resonators   may be taken to be the isotropic form \eqref{eq:dispeqdipoleleadinginpt2}  (which   in fact holds for all values of $\theta_0$) over the entirety of the Brillouin zone. The effectiveness of this approximation is examined numerically in Section \ref{sec:numerics} below.

\section{Helmholtz resonance (cut-off) condition} \label{sec:cutoff}
As determined in Part I and from the denominator of \eqref{eq:dispeqdipoleleadinginpt2} above, the Helmholtz resonance condition in the {\it specially-scaled} limit is given by $1-2 \dbarfeps \approx 0$ and requires careful examination, as such resonators  are capable of    achieving very low-frequency resonances. Accordingly,   we return to the asymptotic form of $\dbarheps$ in \eqref{eq:hepsasybigasp} and subsequently write $1-2 \dbarfeps \approx 0$  in the form
\begin{equation}
\label{eq:rescondhelm}
  \frac{1}{8} + \log\left(\frac{ \pi b}{2\varepsilon}\right)  
 - \left[   \frac{1}{  a^2 } -    \frac{ \emm \pi  }{\varepsilon} - \frac{17}{8} -  \log\left(\frac{\pi a}{8\varepsilon}\right)    \right]\left[  1 + \frac{4\varepsilon \emm}{\pi} \left(   \frac{1}{  a^2 }        \right) \right]^{-1} \approx 0,
\end{equation} 
where we remark that $\tau_5$ in \eqref{eq:listotaus} is taken to $O(1)$ in the numerator and to leading order in the denominator. For the purposes of analysis, 
we do not advise   solving   \eqref{eq:rescondhelm} above  to determine the conditions for resonance, as it is   unclear which terms play a leading role in the small $\varepsilon$ limit;   we   seek the preferred scalings between $\varepsilon$, $a$ (or $b$), and $\emm$ that give the lowest frequency resonance.
  
Therefore, it is helpful to introduce the scalings  $\emm = \kappa_\emm \varepsilon^\mu$ (i.e.,  where $0<\mu<1$, $\varepsilon\rightarrow 0$, and $\kappa_\emm = O(1)$), and  $a = \kappa_a \varepsilon^\gamma$ (i.e.,   where $0<\gamma<1$, $\varepsilon\rightarrow 0$, and $\kappa_a = O(1)$) which admits
\begin{equation}
\label{eq:fepscondalphabeta}
        \frac{1}{8} + \log\left(\frac{ \pi \left[ \kappa_a \varepsilon^{\gamma-1}+2\kappa_\emm \varepsilon^{\mu-1} \right] }{2 }\right)  
 - \dfrac{\left[   \dfrac{1}{   \kappa_a^2 } \varepsilon^{-2\gamma} -    \kappa_\emm \varepsilon^{\mu-1} \pi   - \dfrac{17}{8} -  \log\left(\dfrac{\pi  \kappa_a \varepsilon^{\gamma-1}}{8 }\right)    \right] }{\left[  1 + \dfrac{4  \kappa_\emm }{\pi \kappa_a^2}     \varepsilon^{1 + \mu - 2\gamma} \right]} \approx 0.
\end{equation}
The radii $a  = \kappa_a\varepsilon^\gamma$ and 
$
b = \kappa_a\varepsilon^\gamma+2\kappa_\emm\varepsilon^\mu
$,
must be of the same order, requiring $\mu\geq\gamma$. We now examine, in some detail, the different dominant balance scalings that are possible for the representation \eqref{eq:fepscondalphabeta}.

\subsubsection{Dominant balance in all three numerator terms}
From the form \eqref{eq:fepscondalphabeta}, we take the scaling $1+\mu - 2\gamma>0$, where to ensure that the $O( \varepsilon^{-2\gamma})$, $ O(\varepsilon^{\mu-1})$, and $O(1)$ terms balance in the numerator, we require $\gamma = 0$ and $\mu = 1$. Accordingly,   the   Helmholtz resonance condition   then takes the form
\begin{subequations}
\begin{equation}
\label{eq:fepscondalphabetadombal}
     \frac{9}{4} + 2\log\left(\frac{ \pi   \kappa_a   }{4  \varepsilon  }\right)  
   \approx \left[   \frac{1}{   \kappa_a^2 }  -   \pi   \kappa_\emm         \right] ,
\end{equation}
as $\varepsilon\to 0$. Expressing the above in   dimensional terms we find
\begin{equation}
\label{eq:rescondspecscaled1}
k_\mathrm{max} \approx \frac{1}{\bar{a}} \sqrt{ \dfrac{1}{ 9/4  + 2 \log \left(  \pi \bar{a} /( 4 \bar{\ell })\right) +  \pi \bar{\emm} / \bar{\ell} }	},
\end{equation}
\end{subequations}
as the Helmholtz resonance condition under this dominant balance scaling.

\subsubsection{Dominant balance in numerator pairs} \label{sec:dombalnumpairs}
Another   relationship emerges   by considering   \eqref{eq:fepscondalphabeta} in the $1+\mu - 2\gamma>0$ regime again, which gives 
\begin{equation}
\label{eq:fepscondalphabeta2}
        \frac{9}{4} + \log\left(\frac{ \pi^2 \kappa_a \varepsilon^{\gamma-1} \left[ \kappa_a \varepsilon^{\gamma-1}+2\kappa_\emm \varepsilon^{\mu-1} \right] }{16 }\right)  
 \approx \left[   \frac{1}{   \kappa_a^2 } \varepsilon^{-2\gamma} -  \pi  \kappa_\emm \varepsilon^{\mu-1}   \right]        ,
\end{equation}
where balance on the right-hand side is achieved with the scaling $\gamma = \frac{1}{2}(1-\mu)$, which, as $\mu \geq \gamma$, means that $\mu \geq 1/3$. Turning to the     logarithmic   argument we see that for $\mu>1/3$, the   resonance condition \eqref{eq:fepscondalphabeta2} takes the form
 \begin{subequations}
\label{eq:hrcpartreduced1}
\begin{equation}
\label{eq:fepscondalphabeta3a}
        \frac{9}{4} + \log\left(\frac{ \pi^2 \kappa_a^2 \varepsilon^{-1-\mu }   }{16 }\right)  
 \approx \left[   \frac{1}{   \kappa_a^2 }   -    \pi  \kappa_\emm  \right]  \varepsilon^{\mu-1}      ,
\end{equation}
whereas for $\mu = 1/3$ we observe
\begin{equation}
\label{eq:fepscondalphabeta3zero}
        \frac{9}{4} + \log\left(\frac{ \pi^2 \kappa_a \varepsilon^{-4/3} \left[ \kappa_a  +2\kappa_\emm  \right] }{16 }\right)  
 \approx \left[   \frac{1}{   \kappa_a^2 }   -    \pi  \kappa_\emm  \right]  \varepsilon^{-2/3}      .
\end{equation}
\end{subequations}
Expressing these resonance conditions in terms of dimensional parameters we again find for $\mu>1/3$ the form in \eqref{eq:rescondspecscaled1}, and for $\mu = 1/3$ the slightly different form
\begin{equation}
\label{eq:rescondspecscaled2}
k_\mathrm{max} \approx \frac{1}{\bar{a}} \sqrt{ \dfrac{1}{ 9/4  +   \log \left(  \pi^2 \bar{a} \bar{b} /( 16 \bar{\ell }^2)\right) +  \pi \bar{\emm} / \bar{\ell} }	},
\end{equation}
with numerical investigations suggesting that only minor differences are found between     \eqref{eq:rescondspecscaled1} and  \eqref{eq:rescondspecscaled2}, since $a$ and $b$ must be   the same order. For reference, we remark that under the dominant balance scaling $\mu + 1 -2\gamma>0$,  we may write
\begin{equation}
\label{eq:fepsspecscaled}
\dbarfeps \approx   \frac{b^2 }{ 2}\left[ \frac{1}{a^2} + \frac{1}{b^2} - \frac{9}{4}  - \frac{\pi \emm  }{\varepsilon} -   \log\left( \frac{\pi^2 a b}{16 \varepsilon^2}\right) \right].
\end{equation}

\section{Homogenisation of Helmholtz resonator arrays}\label{sec:homog}
\subsection{Classical homogenisation results for   arrays of ideal cylinders } \label{sec:classicalcylshomog}
For an isotropic fluid medium  of density $\rho$ and bulk modulus $B$, structured with   a two-dimensional array of isotropic fluid {\it cylinders}, of density $\rho_\rmc$ and bulk modulus $B_c$,   the effective density and bulk modulus are  given explicitly in the quasistatic limit  by \cite{torrent2006effective,martin2010estimating}
\begin{equation}
\label{eq:eff2phase}
\rho_\mathrm{eff}^{-1}= \frac{1}{\rho} \frac{\rho_c(1-f) + \rho(1+f)}{\rho_c(1+f) + \rho(1-f)},\quad \mbox{and}\quad 
B_\mathrm{eff}^{-1} = \frac{1-f}{B} + \frac{f}{B_c},
\end{equation}
where $f=\pi b^2/d^2$, as defined previously, denotes the filling fraction.
Subsequently,   results corresponding to  a two-dimensional array of cylinders with     Neumann boundary conditions on the walls take the form
\begin{subequations}
\begin{align}
\label{eq:cylneueff}
\rho_\mathrm{eff}^{-1} =  \frac{1-f}{\rho(1+f)}, \quad \mbox{and}\quad B_\mathrm{eff}^{-1} = \frac{1-f}{B},
\end{align}
which are obtained via the limits $B_c \rightarrow \infty$ and $\rho_c \to \infty$ in \eqref{eq:eff2phase} above.  Substituting   \eqref{eq:cylneueff} into the (dimensional) dispersion equation for plane waves in an unbounded isotropic medium 
\begin{equation}
\label{eq:freespacedispreleff}
 \bar{k}_\rmB \, \rho_\mathrm{eff}^{-1} \, \bar{k}_\rmB - \omega^2 B_\mathrm{eff}^{-1}  = 0,
\end{equation}
 we obtain the   (dimensionless) dispersion equation for an array of Neumann cylinders
\begin{equation}
k_\rmB^2 = 1+f,
\end{equation}
\end{subequations}
  as seen from our analysis in Part I (Eq.~(4.23)) and from  Movchan {\it et. al} \cite[Eq. (3.158)]{movchan2002asymptotic}, and where $\bar{\bfk}_\rmB = k \bfk_\rmB$ is the nondimensional Bloch vector. Similarly, if we   return  to the two-phase fluid array results \eqref{eq:eff2phase} and take the limits $B_c \rightarrow B$ and $\rho_c \to \infty$ we obtain
\begin{subequations}
\begin{align}
\label{eq:beffrhoeffstrlimit}
\rho_\mathrm{eff}^{-1} =  \frac{1-f}{\rho(1+f)}, \quad \mbox{and}\quad
B_\mathrm{eff}^{-1} = \frac{1}{B},
\end{align}
which on substitution in \eqref{eq:freespacedispreleff} returns the dispersion relation
\begin{equation}
\label{eq:dispeq0order}
k_\rmB^2 = \frac{1+f}{1-f},
\end{equation}
\end{subequations}
and is identical to the lowest-order (isotropic) approximation for an array of thin-walled Helmholtz resonators at quasi-static frequencies (i.e., $\omega\rightarrow 0$), as given in Part I and in Llewellyn--Smith \cite{llewellyn2010split}. That is,   the lowest-order (isotropic) approximation for the  dispersion equation  of a Helmholtz resonator array \eqref{eq:dispeq0order}  {\it at very low  frequencies} is  indistinguishable from  an array of fluid cylinders   that do not possess a contrast in bulk modulus but are much denser than the background fluid.  On comparing   \eqref{eq:cylneueff} and \eqref{eq:beffrhoeffstrlimit} we see that  the impact of introducing a small gap into the wall of a perfect Neumann cylinder    has the leading-order effect of modifying the effective bulk modulus  but not   the effective density. As such,  we may   expect to recover the same   expressions for the bulk modulus and density in \eqref{eq:beffrhoeffstrlimit} at low frequencies from any isotropic  descriptions for resonator arrays.

\subsection{Isotropic descriptions for thin-walled Helmholtz resonator arrays} \label{sec:isothindescript}
Having discussed the behaviour of the isotropic description at zero frequency, we now restate    the   result obtained in Part I (Eq.~(4.22))
 \begin{equation}
\label{eq:dispeqdipoleleadinginpt3}
k_\rmB^2 = \frac{1+f}{1-f} \left(1 -  \frac{   2 f (1 - f_\varepsilon )}{1 - 2f_\varepsilon} \right),
\end{equation} 
which is valid for frequencies across the range of the  first band surface, where
 \begin{equation}
 \label{eq:fepsasymptpt20}
f_\varepsilon   \sim 1 - b^2/8+ b^2\log\left(\theta_\mathrm{ap}/2  \right),
\end{equation}
and  $\theta_\mathrm{ap}$ denotes the aperture half-width angle.  Here, we emphasise that there are infinitely many ways in which this dispersion equation may be decomposed into the form \eqref{eq:freespacedispreleff}. For example,  
we may extract from \eqref{eq:dispeqdipoleleadinginpt3} the  response functions  
\begin{equation}
\label{eq:combo2}
\rho_\mathrm{eff}^{-1} =  \frac{(1-f)}{\rho(1+f)},  \quad \mbox {and} \quad
B_\mathrm{eff}^{-1}  =  \frac{1}{B} \frac{(1-2f_\varepsilon - 2f(1-f_\varepsilon))}{(1-2f_\varepsilon)},
\end{equation}
as   candidates for the behaviour of the resonator array, within an isotropic approximation. Such expressions are consistent with the two-phase fluid results \eqref{eq:beffrhoeffstrlimit} as $\omega\rightarrow 0$,   vanishing filling fraction results $f\rightarrow 0$, and the expressions for closed cylinders \eqref{eq:cylneueff}
    as the aperture is closed (i.e., as $f_\varepsilon \rightarrow \infty$). The absence of dispersion in the density response is also consistent with existing literature on the topic \cite{schweizer2017resonance,lamacz2016effective}. Thus, we consider \eqref{eq:combo2} as the effective (homogenised) quantities of the medium for frequencies spanning the first band surface, within an isotropic approximation.

\subsection{Anisotropic descriptions for thin-walled Helmholtz resonator arrays} \label{sec:anisothinsect}
As established in Part I, thin-walled resonators generally exhibit   strong anisotropy at low frequencies, and so   isotropic descriptions are   insufficient to accurately describe the first band surface. Accordingly, we restate the  anisotropic  result for the first spectral band from Part I (Eq.~(4.29a)):
\begin{equation}
\label{eq:dispeqfirstorder}
k_\rmB^2 = \frac{(f+1) \left[b^2 f  (2 f-1) +   f_\varepsilon   (f+1) (2 f_\varepsilon f-2 f_\varepsilon-2 f+1)\right]}{b^2 f  \cos (2 \left[\theta_0-\theta_\rmB\right])+b^2 f^2+  f_\varepsilon (2 f_\varepsilon-1) \left(f^2-1\right)},
\end{equation}
and emphasise once more  that the assignment of the effective density and bulk modulus is non-unique in the dispersion equation for plane waves in an anisotropic medium  \cite{norris2015acoustic}
\begin{equation}
\label{eq:freespacedispreleffaniso}
 \bar{k}_{\rmB i} \, (\rho_\mathrm{eff}^{-1})_{ij}  \, \bar{k}_{\rmB j} - \omega^2 B_\mathrm{eff}^{-1}  = 0.
\end{equation}
For example,   we may propose the candidate forms
\begin{subequations}
\label{eq:anisorhoBrespfuncs}
\begin{align}
\boldsymbol{\rho}^{-1}_\mathrm{eff} &= 
\frac{(1-f)}{\rho(1+f)}
\left[
\begin{array}{c|c}
1 - b^2 f\left[1 - \cos(2\theta_0)\right]/ \mathcal{H} & b^2 f \sin (2 \theta_0)/ \mathcal{H}\\ \hline
b^2 f \sin (2 \theta_0)/ \mathcal{H} &1 - b^2 f\left[1 + \cos(2\theta_0)\right]  / \mathcal{H}
\end{array}
\right]
\\
B_\eff^{-1} &= \frac{(1-f)}{B \mathcal{H}}   \left[  b^2 f (2 f-1)+(f+1) f_\varepsilon (2 (f-1) f_\varepsilon-2 f+1)\right] ,
\end{align}
\end{subequations}
where 
\begin{equation}
\label{eq:helmresfreqaniso}
\mathcal{H} = b^2 f^2+b^2 f+ (f^2-1 ) f_\varepsilon (2 f_\varepsilon-1) =0.
\end{equation}
These expressions are consistent with both the  isotropic results \eqref{eq:combo2} and the two-phase fluid results \eqref{eq:beffrhoeffstrlimit} as $\omega\rightarrow 0$, as well as with the closed aperture (Neumann cylinder) results \eqref{eq:cylneueff} as $f_\varepsilon \rightarrow \infty$ and the vanishing fill fraction limit $f\rightarrow 0$.  Note that the expression $\mathcal{H}$  is the anisotropic approximation to the Helmholtz resonance condition      $\det  \{ (\rho_\mathrm{eff}^{-1})_{ij}  \}=0$.

\subsection{Descriptions for moderately thick-walled Helmholtz resonator arrays }\label{sec:isothickmodsect}
For   moderately thick-walled resonators, the analysis proceeds as in   Section \ref{sec:homog}\ref{sec:isothindescript} and   \ref{sec:homog}\ref{sec:anisothinsect} above but with the replacement $f_\varepsilon \mapsto \check{f}_\varepsilon$, where the definition for $\check{f}_\varepsilon$ is given in Part I (Eq.~(6.14)).

\subsection{Descriptions for specially-scaled   Helmholtz resonator arrays} \label{sec:isosstwhelmres}
To obtain an isotropic description for the {\it specially-scaled} thick-walled resonator configuration discussed in this Part II, the analysis proceeds as in  Section \ref{sec:homog}\ref{sec:isothindescript}  with the replacement $f_\varepsilon \mapsto \dbarfeps$  where $\dbarfeps$ is defined in \eqref{eq:fepsspecscaled}. As shown   in Section \ref{eq:asyregsystem}\ref{sec:disprelsstwhelmres} earlier,     anisotropic descriptions derived    from the regularised system for {\it specially-scaled}  resonators are generally    intractable, i.e., see \eqref{eq:disprelscalGY}. That said, it should be possible to obtain an anisotropic description of the first spectral band alone   using \eqref{eq:dispeqfirstorder} with the replacement $f_\varepsilon \mapsto \dbarfeps$.

\section{Numerical Results} \label{sec:numerics}
In this section, we compute a broad selection of band diagrams, comparing results from our regularised system \eqref{eq:dispeqsystemgnptfinal} and   asymptotic dispersion equation   \eqref{eq:dispeqdipoleleadinginpt2} against a full finite-element treatment. We examine the impact of varying the aspect ratio $h = 2\emm/2\ell = (\bar{b} - \bar{a})/2\bar{\ell}$, varying the aperture width $\theta_\mathrm{ap}$, and varying the filling fraction $f$ upon the spectral behaviour of the array, as well as its impact on the Helmholtz resonance frequency. We also evaluate expressions for the effective inverse density and effective inverse Bulk modulus (compressibility) for a selection of thin-walled, moderately thick-walled, and {\it specially-scaled} thick-walled Helmholtz resonator arrays  in Figures \ref{fig:respfuncsthinwall} and \ref{fig:respfuncsthickerwalls}, within the isotropic  \eqref{eq:combo2} and anisotropic  \eqref{eq:anisorhoBrespfuncs} approximations given earlier. For reference we consider wave propagation through air and     take $B   = 0.14183$ MPa and $\rho   =  1.2041$ kg/m$^3$ accordingly.

  \begin{figure}[t]
\centering
\subfloat[Subfigure 6 list of figures text][]{
\includegraphics[width=0.475\textwidth]{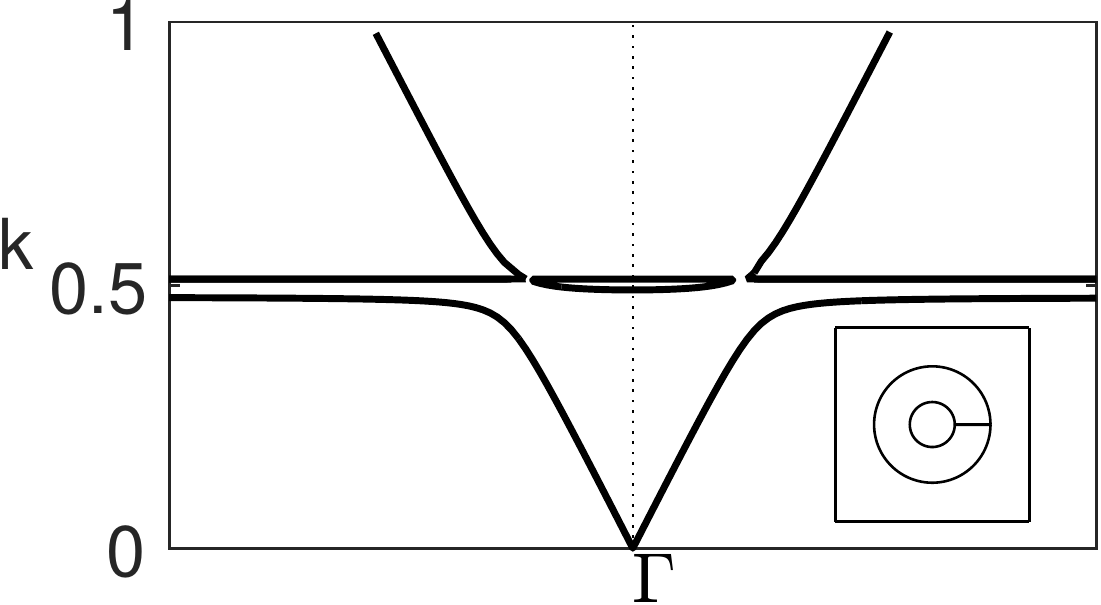}
\label{fig:fold1}}
\subfloat[Subfigure 1 list of figures text][]{
\includegraphics[width=0.475\textwidth]{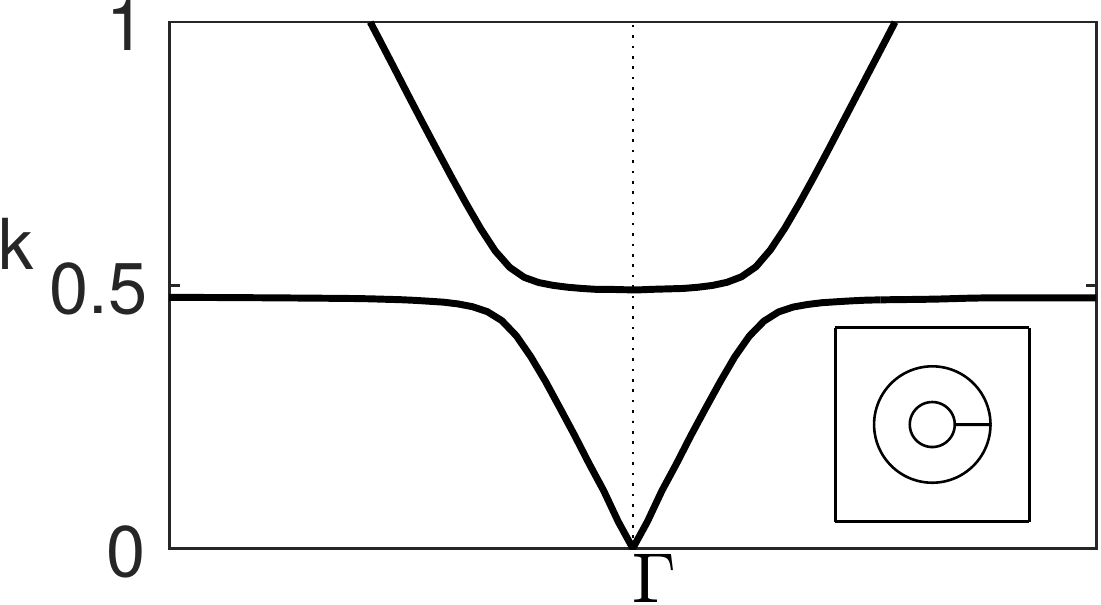}
\label{fig:fold2}}
\caption{Comparison of band diagrams (near the $\Gamma$ point) for a two-dimensional square array of specially-scaled   Helmholtz resonators obtained using multipole methods. Figure \protect\subref{fig:fold1}  shows a spurious result obtained using the system   from Part I  \eqref{eq:dispeqsystemgnptii} with the updated $\dbarheps$   \eqref{eq:hepsbigasp}, and   \protect\subref{fig:fold2}  shows the correct result obtained using the regularised system \eqref{eq:dispeqsystemgnptfinal}. Inset: corresponding fundamental unit cells. Results  are given  for a dipole truncation $L=1$ and truncations $L_{\chi_n} = L_{F_m} = L_{H_\varepsilon}=13$ with $\bar{d} = 1$, $\theta_0 = 0$,   $\bar{b} = 0.3$, $h=100$, and $\theta_\mathrm{ap} = \pi/1024$.}
\label{fig:foldedboth}
\end{figure}

In Figure \ref{fig:foldedboth} we examine the band diagram for a {\it specially-scaled}   resonator near the $\Gamma$   symmetry point, comparing the result obtained from the system in Part I (with the replacement $h_\varepsilon \mapsto \dbarheps$) against that obtained from the regularised system \eqref{eq:dispeqsystemgnptfinal} directly. In the former case,  Figure \ref{fig:fold1}    demonstrates  unexpected spectral behaviour in the form of band folding effects near $k$ values corresponding to $\dbarheps \approx 0$. At the cusps of the folded bands, the group velocity $\partial \omega / \partial \bar{k}_{Bi}$ is in principle infinite \cite{chen2011folded}, although as shown in Figure \ref{fig:fold2}, such features are in fact spurious and are not a feature of the genuine system. Such behaviour demonstrates the need to exercise appropriate caution when calculating band diagrams using multipole methods, although we remark that spurious spectral behaviour can   be overcome using numerical techniques: by searching  for both the zero determinant and  vanishing minimum singular value of the matrix system and only considering those values which satisfy both measures.

  \begin{figure}[t]
\centering
\subfloat[Subfigure 6 list of figures text][]{
\includegraphics[width=0.475\textwidth]{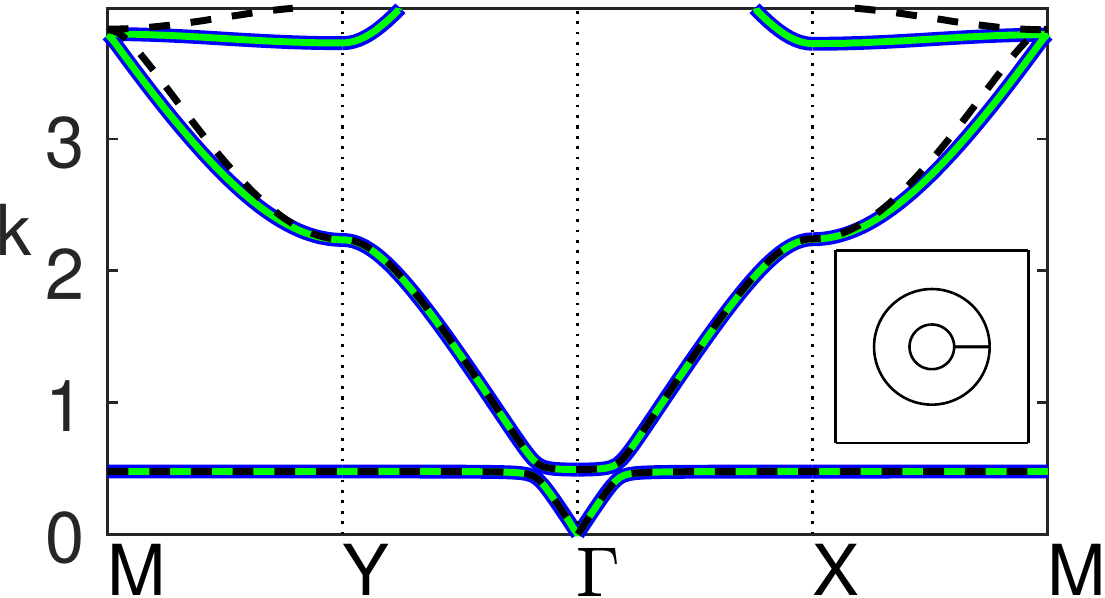}
\label{fig:conv1}}
\subfloat[Subfigure 1 list of figures text][]{
\includegraphics[width=0.475\textwidth]{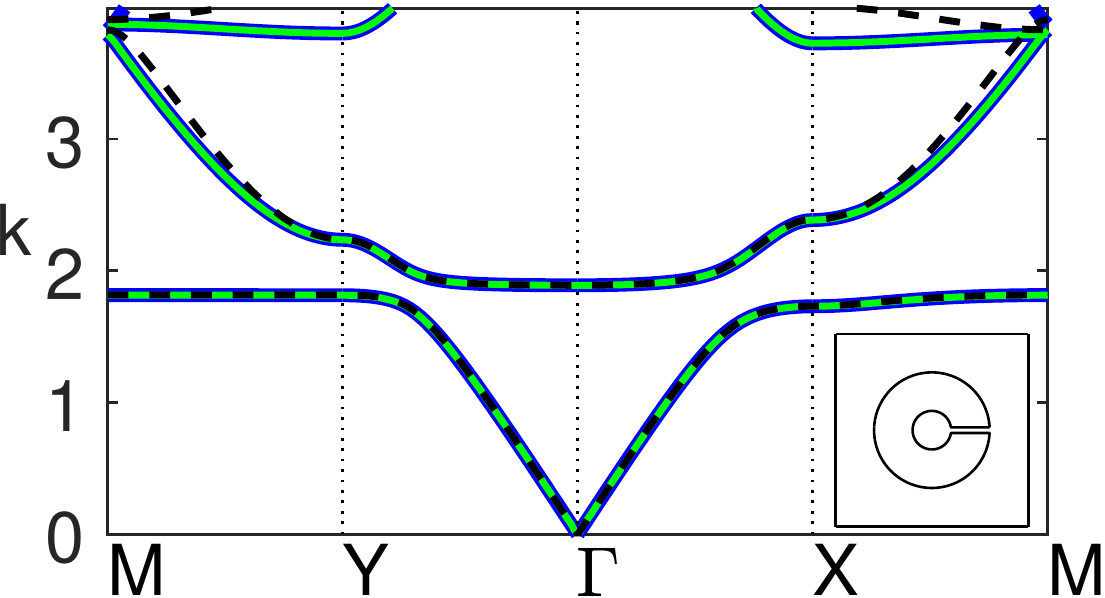}
\label{fig:conv2}}
\caption{Band diagrams for  two-dimensional square arrays of specially-scaled   Helmholtz resonators  comparing the finite-element solution (blue curves) against  results obtained using the regularised system \eqref{eq:dispeqsystemgnptfinal} for dipole $L=1$ (dashed black curves) and quadrupole $L=3$ (green curves)   truncations.  Inset: corresponding fundamental unit cells. Figure  \protect\subref{fig:fold1} corresponds to  $h=100$  ($\bar{a} \approx 0.116$) and $\theta_\mathrm{ap} = \pi/1024$, and  \protect\subref{fig:fold2} corresponds to $\bar{a} = 0.1$ 
and  $\theta_\mathrm{ap} = \pi/64$. In both figures, we use the truncations $L_{\chi_n} = L_{F_m} = L_{H_\varepsilon}=13$ with $\bar{d} = 1$, $\theta_0 = 0$,  and $\bar{b} = 0.3$.
  }
\label{fig:convergenceboth}
\end{figure}

\begin{figure}[t]
\centering
\subfloat[Subfigure 6 list of figures text][]{
\includegraphics[width=0.475\textwidth]{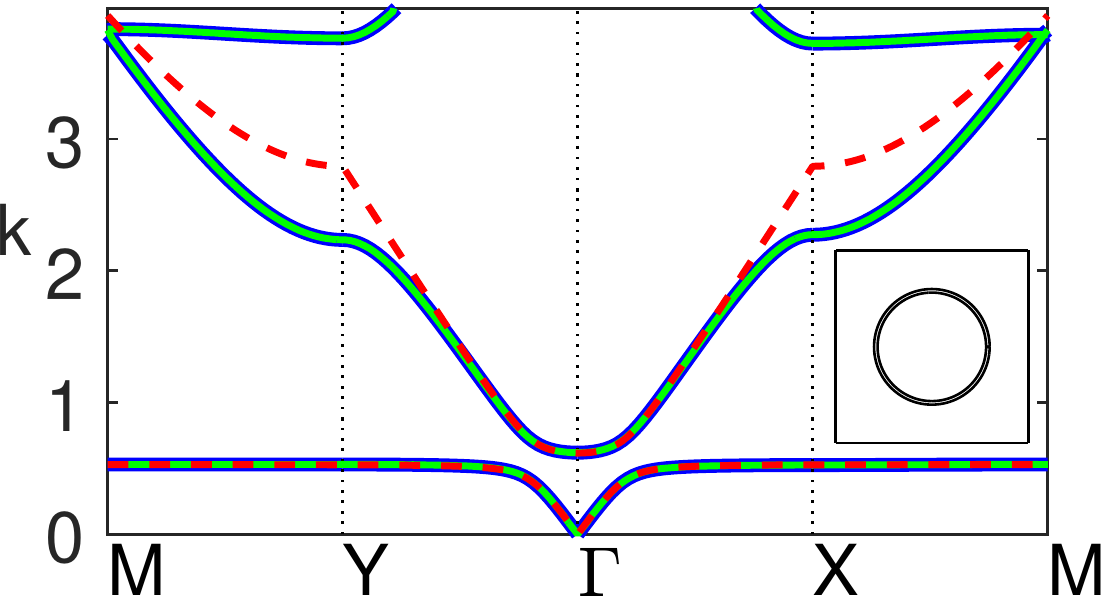}
\label{fig:thickmultipolesubfig1}}
\subfloat[Subfigure 1 list of figures text][]{
\includegraphics[width=0.475\textwidth]{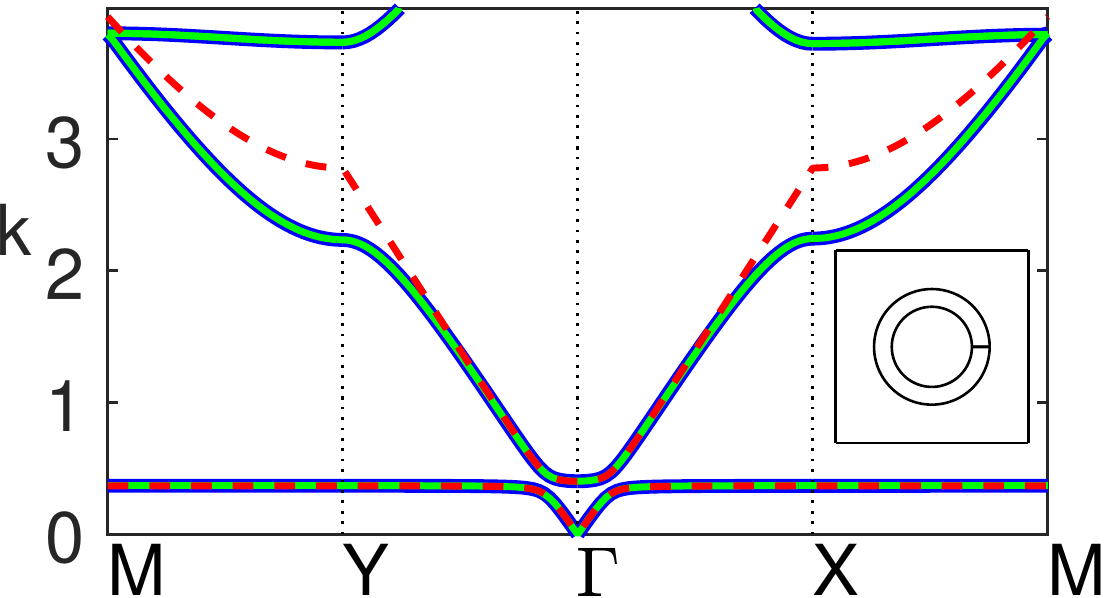}
\label{fig:thickmultipolesubfig2}}\\
\subfloat[Subfigure 2 list of figures text][]{
\includegraphics[width=0.475\textwidth]{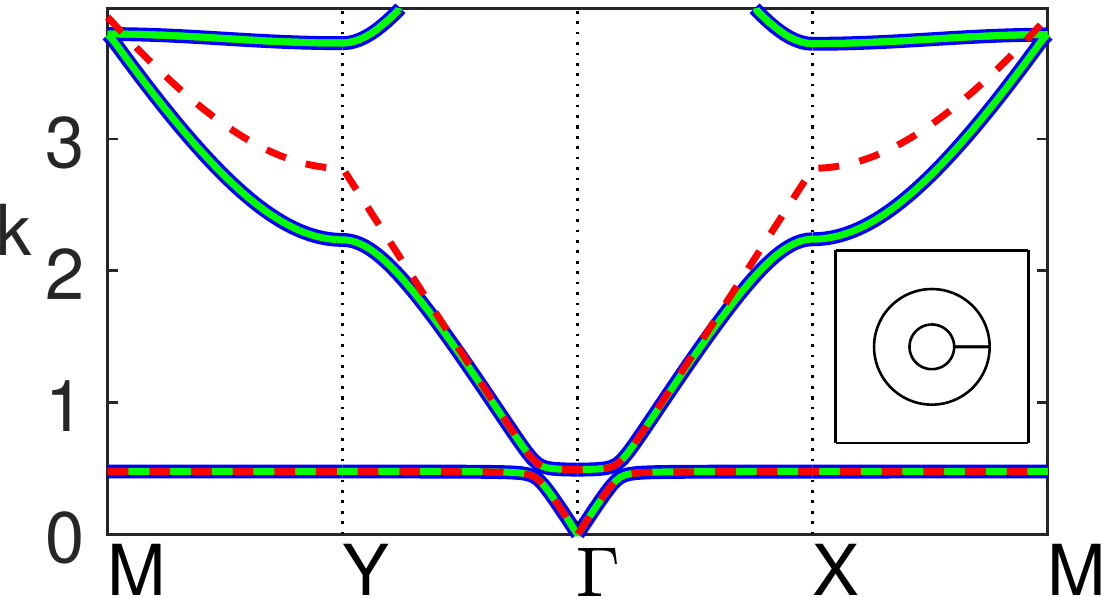}
\label{fig:thickmultipolesubfig3}}
\subfloat[Subfigure 3 list of figures text][]{
\includegraphics[width=0.475\textwidth]{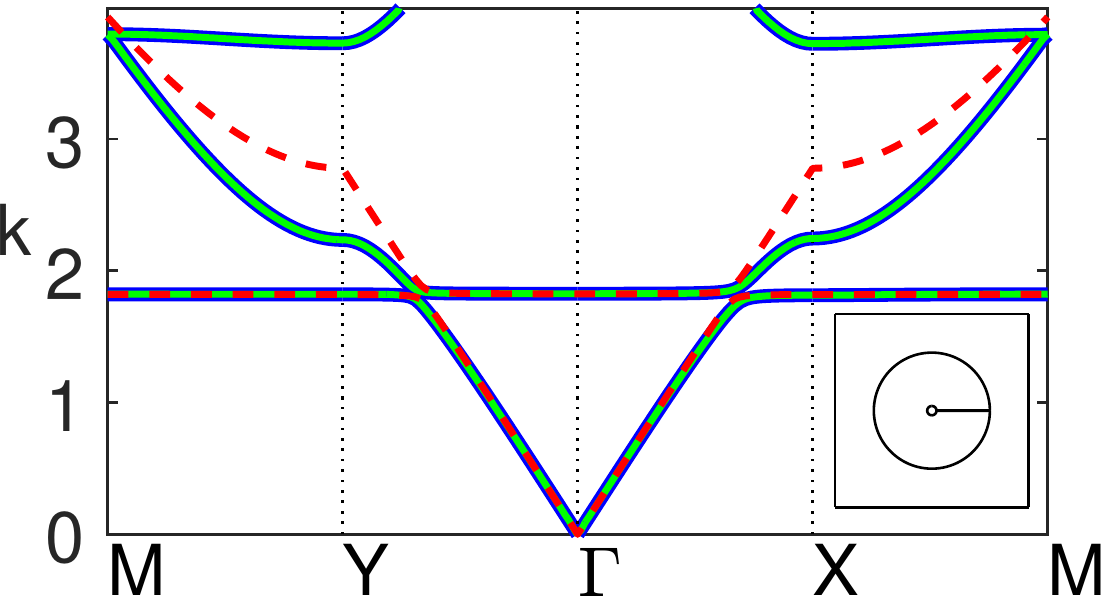}
\label{fig:thickmultipolesubfig4}}
\caption{Band diagrams for a two-dimensional square array of specially-scaled thick-walled Helmholtz resonators  as  the channel aspect ratio $h=\emm/\varepsilon$   is increased:  \protect\subref{fig:thickmultipolesubfig1}  $h=10$, \protect\subref{fig:thickmultipolesubfig2} $h=50$, \protect\subref{fig:thickmultipolesubfig3} $h=100$, and \protect\subref{fig:thickmultipolesubfig4} $h=150$, with fundamental unit cells inset. Multipole results from the regularised system \eqref{eq:dispeqsystemgnptfinal} are given (green lines) for a system truncation $L=3$ and truncations $L_{\chi_n} = L_{F_m} = L_{H_\varepsilon}=13$ with the isotropic approximation \eqref{eq:dispeqdipoleleadinginpt2} (dashed red lines) superposed, in addition to finite-element results (blue lines).  In the above figures we use $\bar{d} = 1$, $\theta_0 = 0$,   $\bar{b} = 0.3$, and $\theta_\mathrm{ap} = \pi/1024$.}
\label{fig:varyingh}
\end{figure}

\begin{figure}[t]
\centering
\subfloat[Subfigure 6 list of figures text][]{
\includegraphics[width=0.475\textwidth]{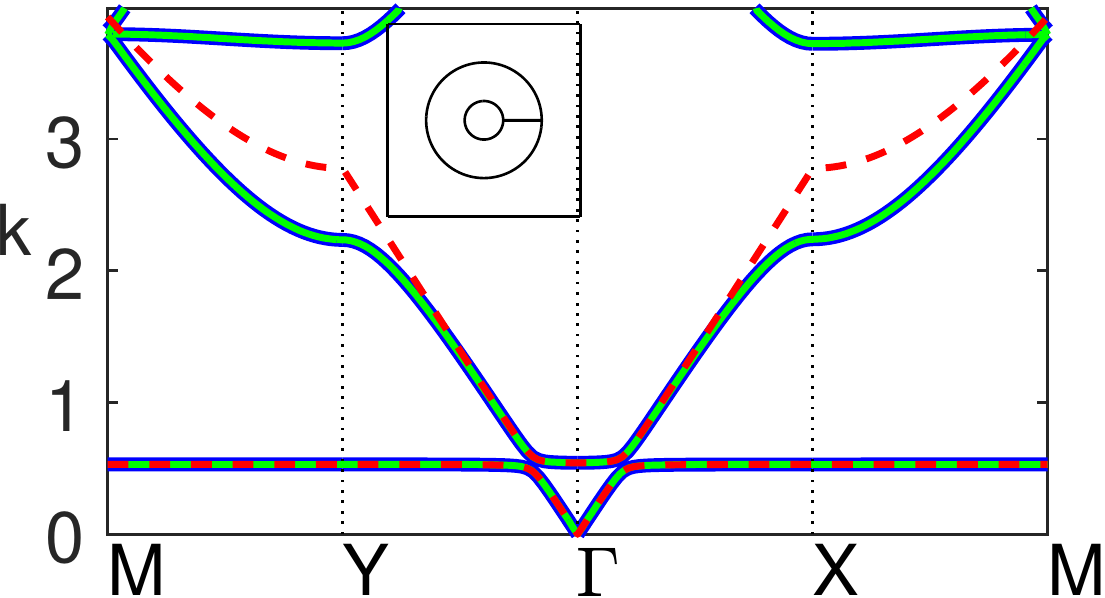}
\label{fig:thicktmultipolesubfig1}}
\subfloat[Subfigure 1 list of figures text][]{
\includegraphics[width=0.475\textwidth]{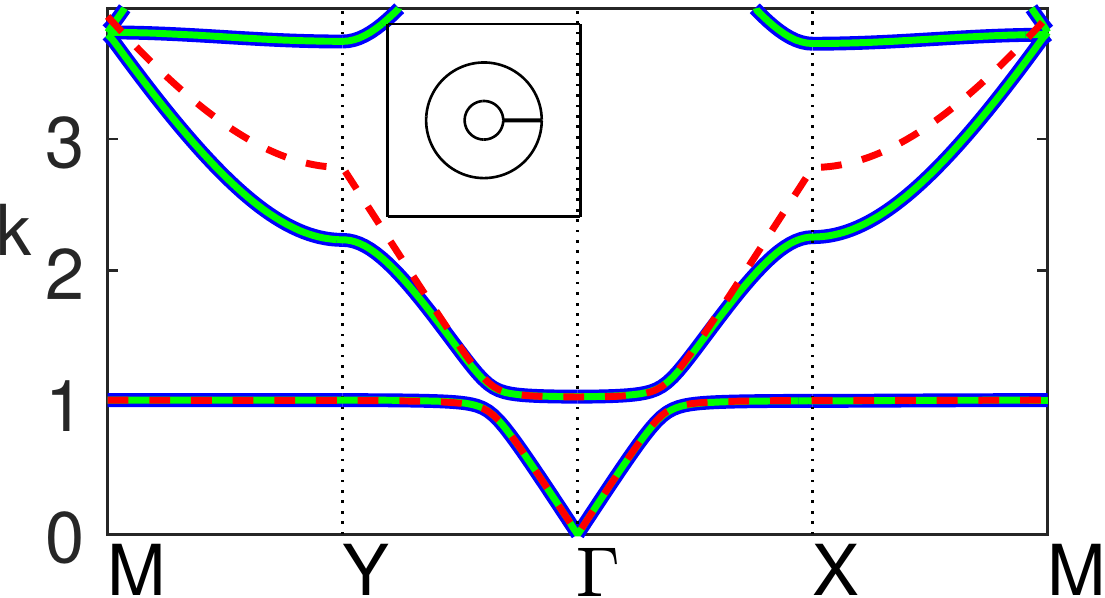}
\label{fig:thicktmultipolesubfig2}}\\
\subfloat[Subfigure 2 list of figures text][]{
\includegraphics[width=0.475\textwidth]{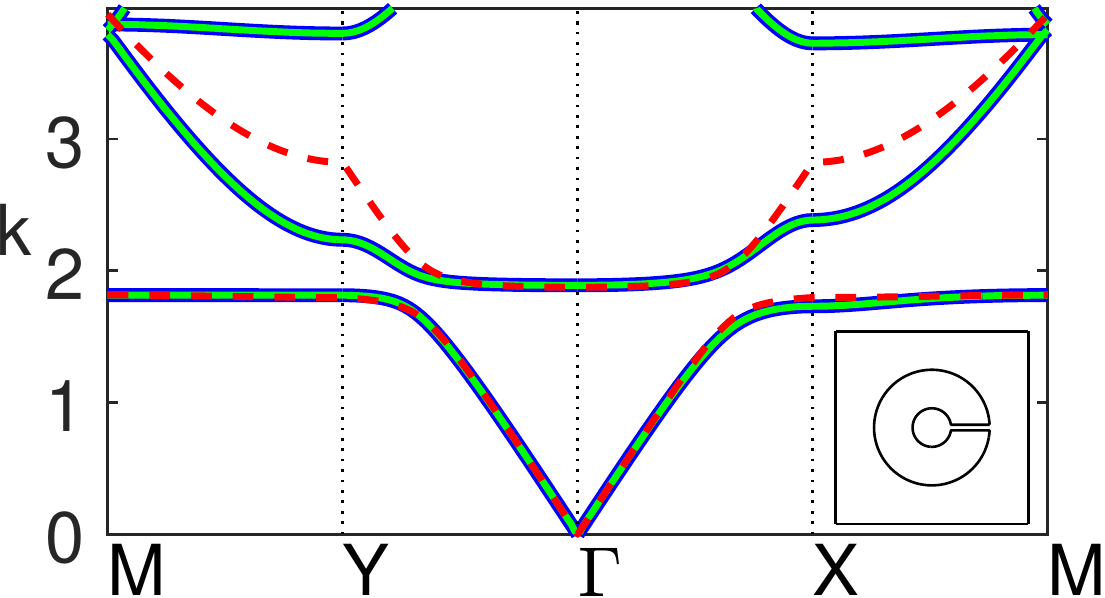}
\label{fig:thicktmultipolesubfig3}}
\subfloat[Subfigure 3 list of figures text][]{
\includegraphics[width=0.475\textwidth]{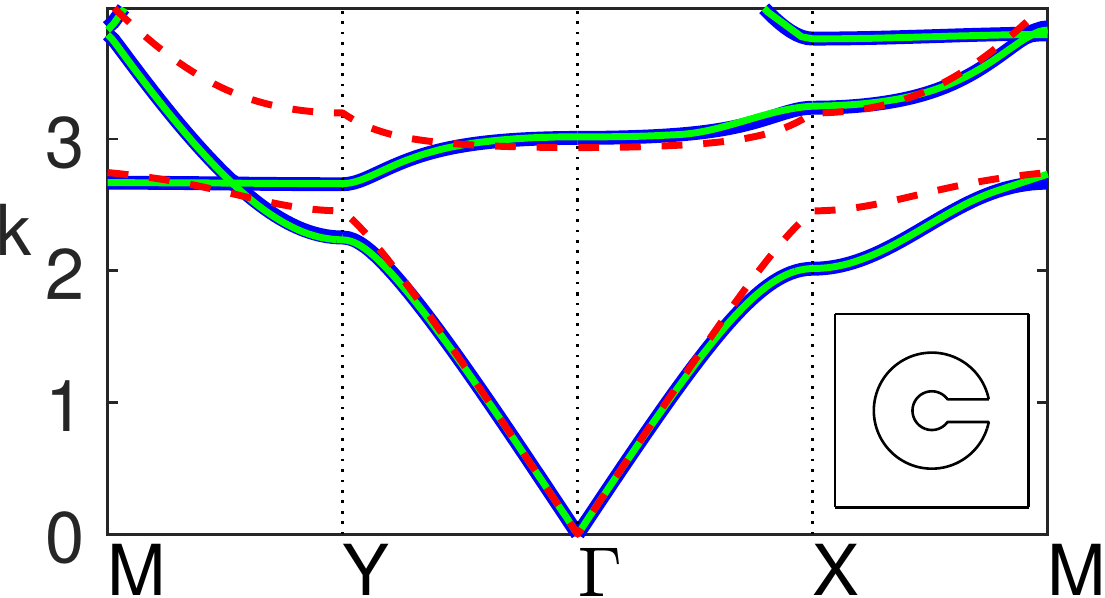}
\label{fig:thicktmultipolesubfig4}}
\caption{Band diagrams for a two-dimensional square array of specially-scaled thick-walled Helmholtz resonators  as  the aperture half-angle $\theta_{\rma\rmp}$   is increased:  \protect\subref{fig:thicktmultipolesubfig1}  $\theta_{\rma\rmp} = \pi/1024$, \protect\subref{fig:thicktmultipolesubfig2} $\theta_{\rma\rmp} = \pi/256$, \protect\subref{fig:thicktmultipolesubfig3} $\theta_{\rma\rmp} = \pi/64$, and \protect\subref{fig:thicktmultipolesubfig4} $\theta_{\rma\rmp} = \pi/16$, with fundamental unit cells inset. Figure legends and truncation values are identical to those in Figure \ref{fig:varyingh}; here   we use $\bar{d} = 1$, $\theta_0 = 0$,   $\bar{b} = 0.3$, and $\bar{a} = 0.1$.}
\label{fig:varyingthetaap}
\end{figure}

\begin{figure}[t]
\centering
\subfloat[Subfigure 6 list of figures text][]{
\includegraphics[width=0.475\textwidth]{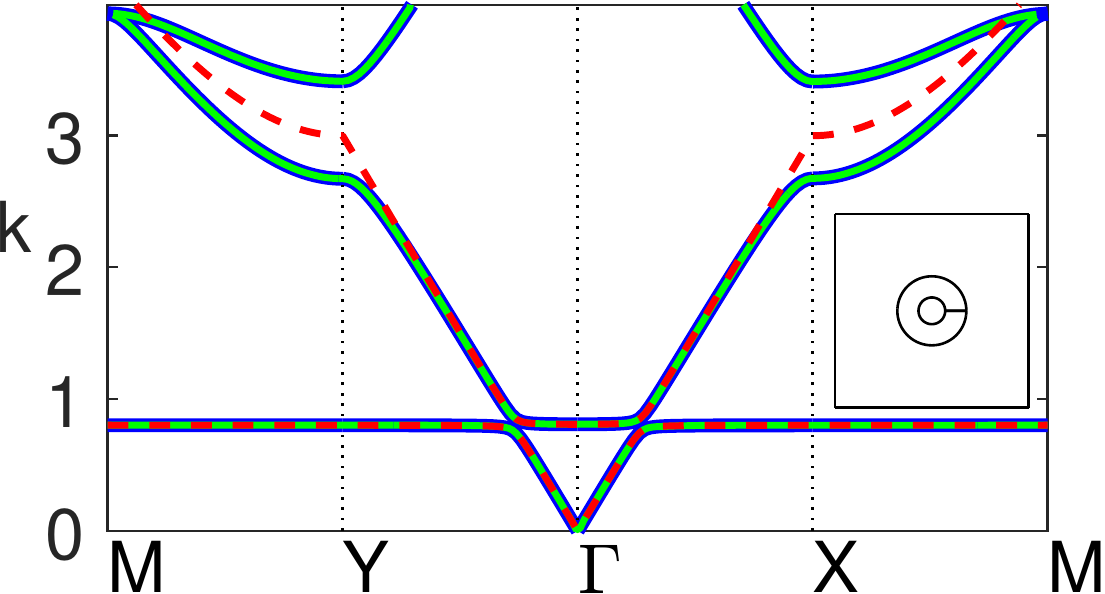}
\label{fig:thickfmultipolesubfig1}}
\subfloat[Subfigure 1 list of figures text][]{
\includegraphics[width=0.475\textwidth]{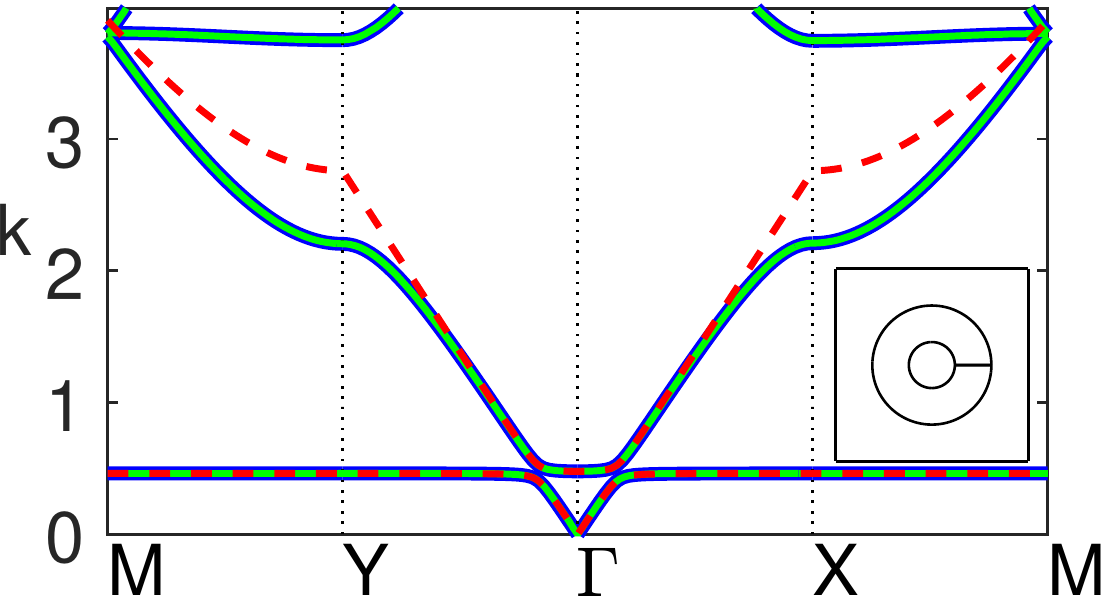}
\label{fig:thickfmultipolesubfig2}}\\
\subfloat[Subfigure 2 list of figures text][]{
\includegraphics[width=0.475\textwidth]{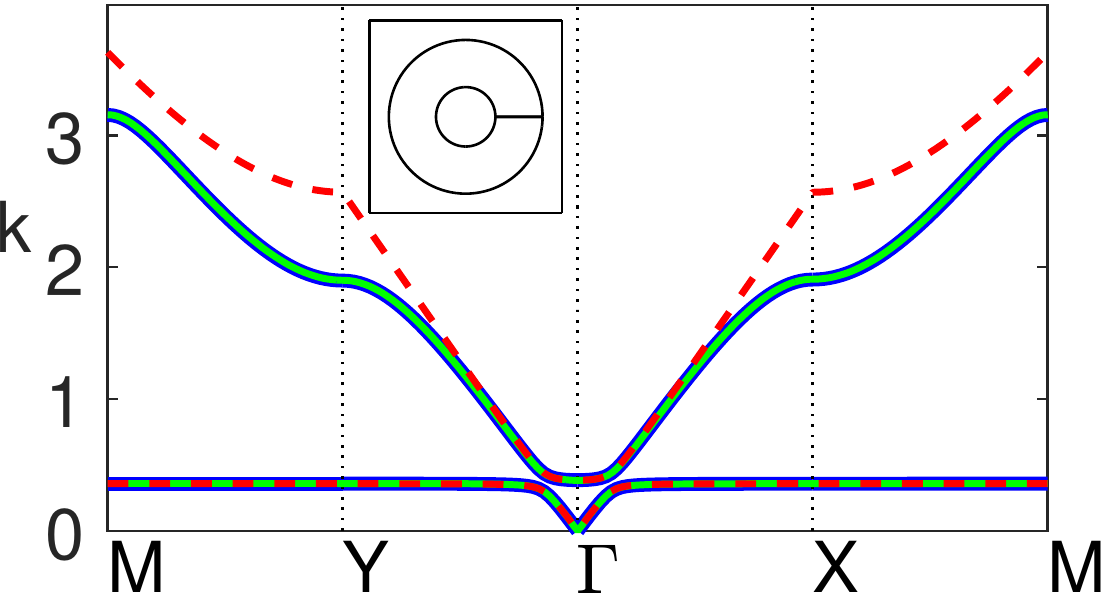}
\label{fig:thickfmultipolesubfig3}}
\subfloat[Subfigure 3 list of figures text][]{
\includegraphics[width=0.475\textwidth]{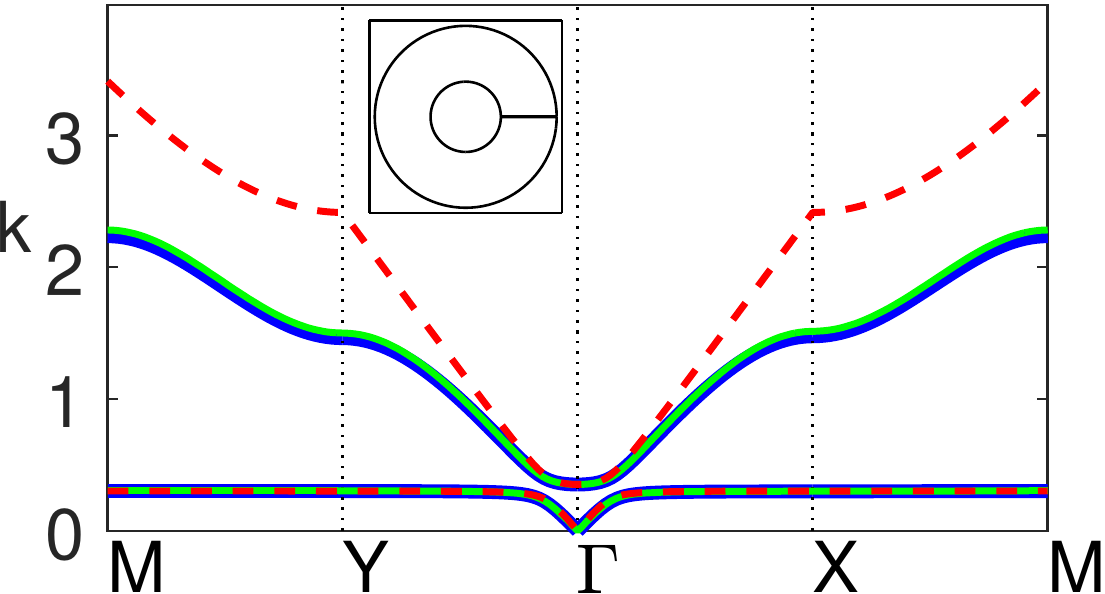}
\label{fig:thickfmultipolesubfig4}}
\caption{Band diagrams for a two-dimensional square array of specially-scaled thick-walled Helmholtz resonators  as  the filling fraction $f$  is increased:  \protect\subref{fig:thickfmultipolesubfig1}  $f=0.1$, \protect\subref{fig:thickfmultipolesubfig2} $f=0.3$, \protect\subref{fig:thickfmultipolesubfig3} $f=0.5$, and \protect\subref{fig:thickfmultipolesubfig4} $f=0.7$, with fundamental unit cells inset.  Figure legends and truncation values are identical to those in Figure \ref{fig:varyingh};
here we use $\bar{d} = 1$, $\theta_0 = 0$,  $h = 100$, and $\theta_\mathrm{ap} = \pi/1024$.}
\label{fig:varyingf}
\end{figure}

\begin{figure}[t]
\centering
\subfloat[Subfigure 6 list of figures text][]{
\includegraphics[width=0.425\textwidth]{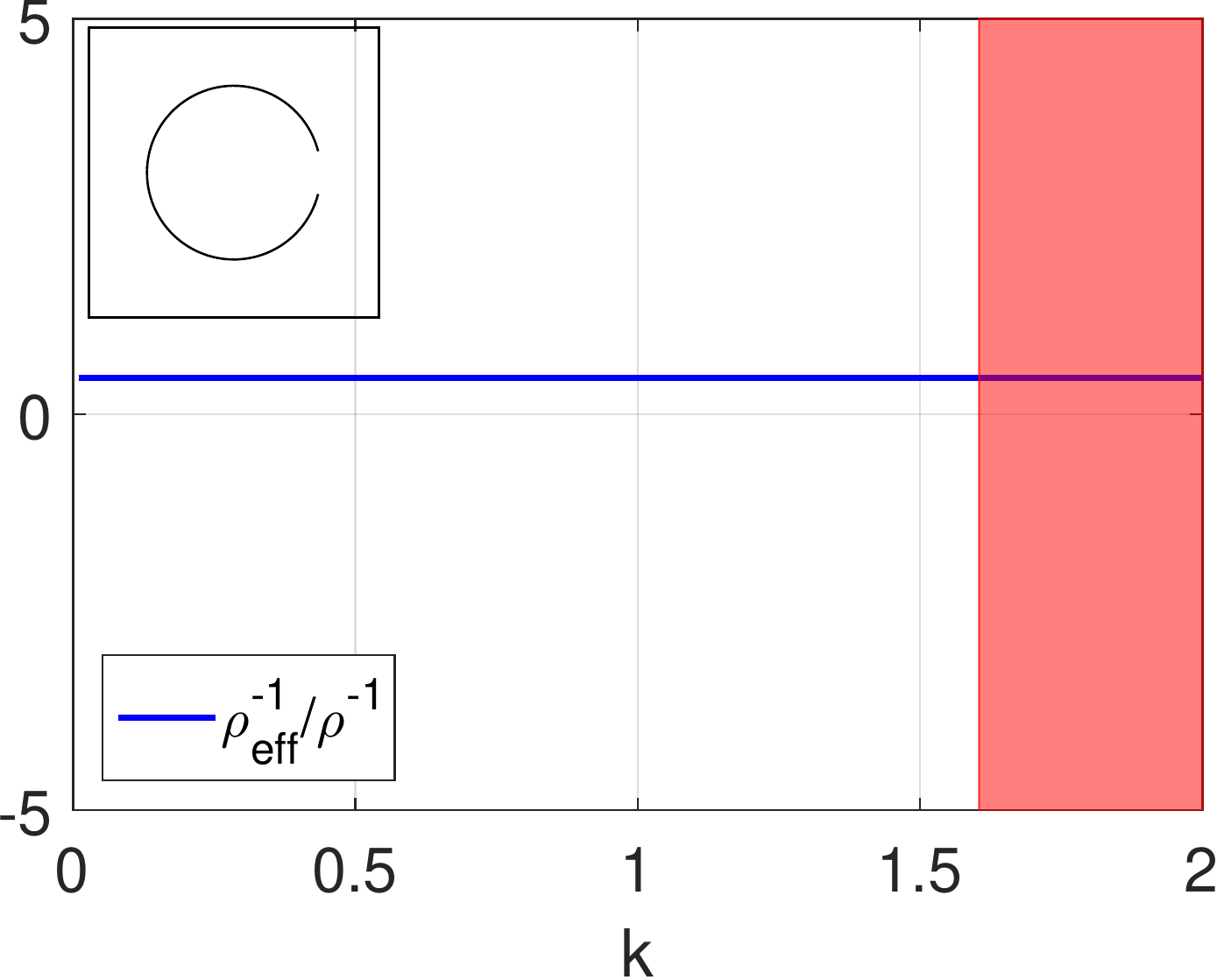}
\label{fig:rhoinvisothin}}
\subfloat[Subfigure 1 list of figures text][]{
\includegraphics[width=0.425\textwidth]{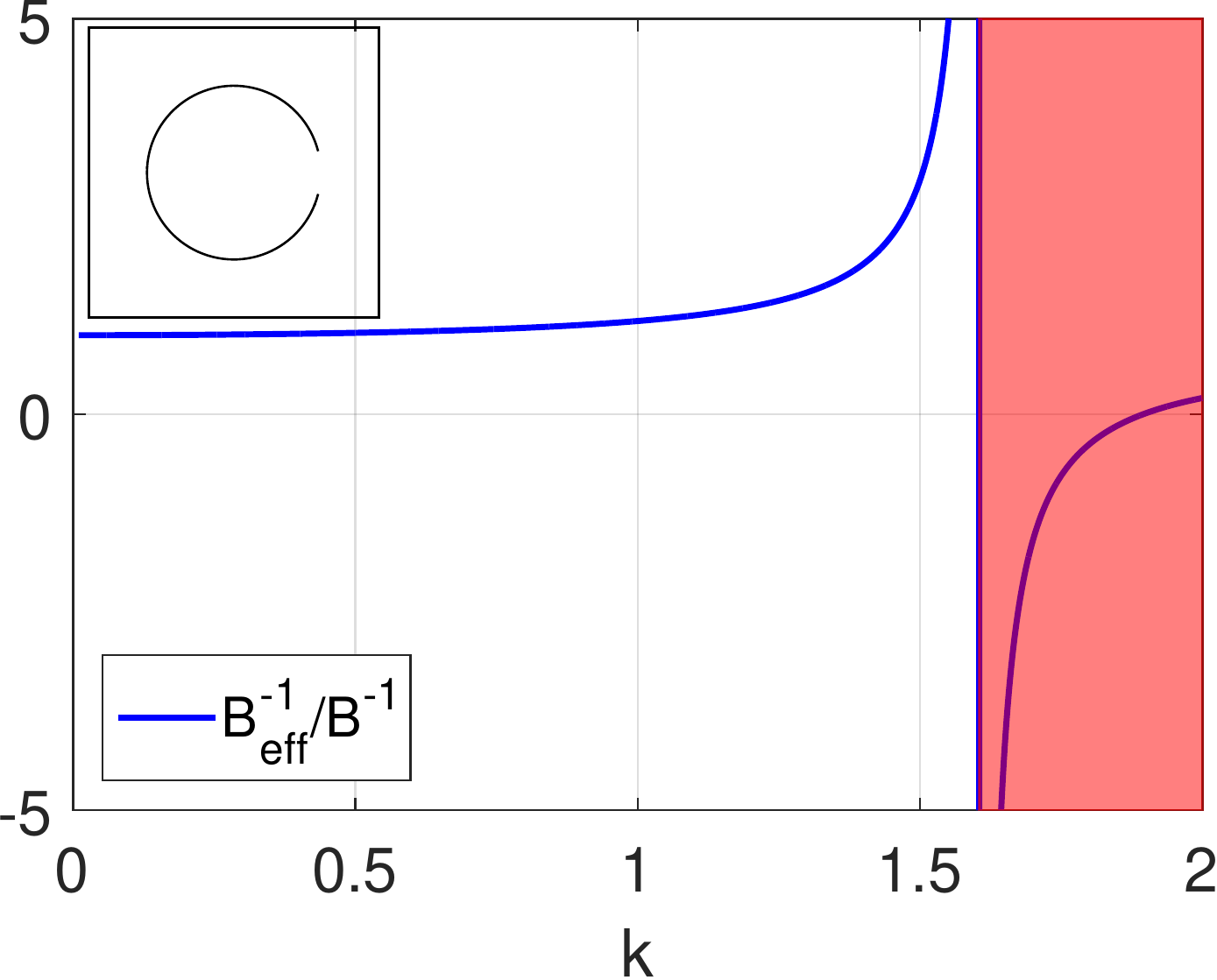}
\label{fig:Binvisothin}}\\
\subfloat[Subfigure 2 list of figures text][]{
\includegraphics[width=0.425\textwidth]{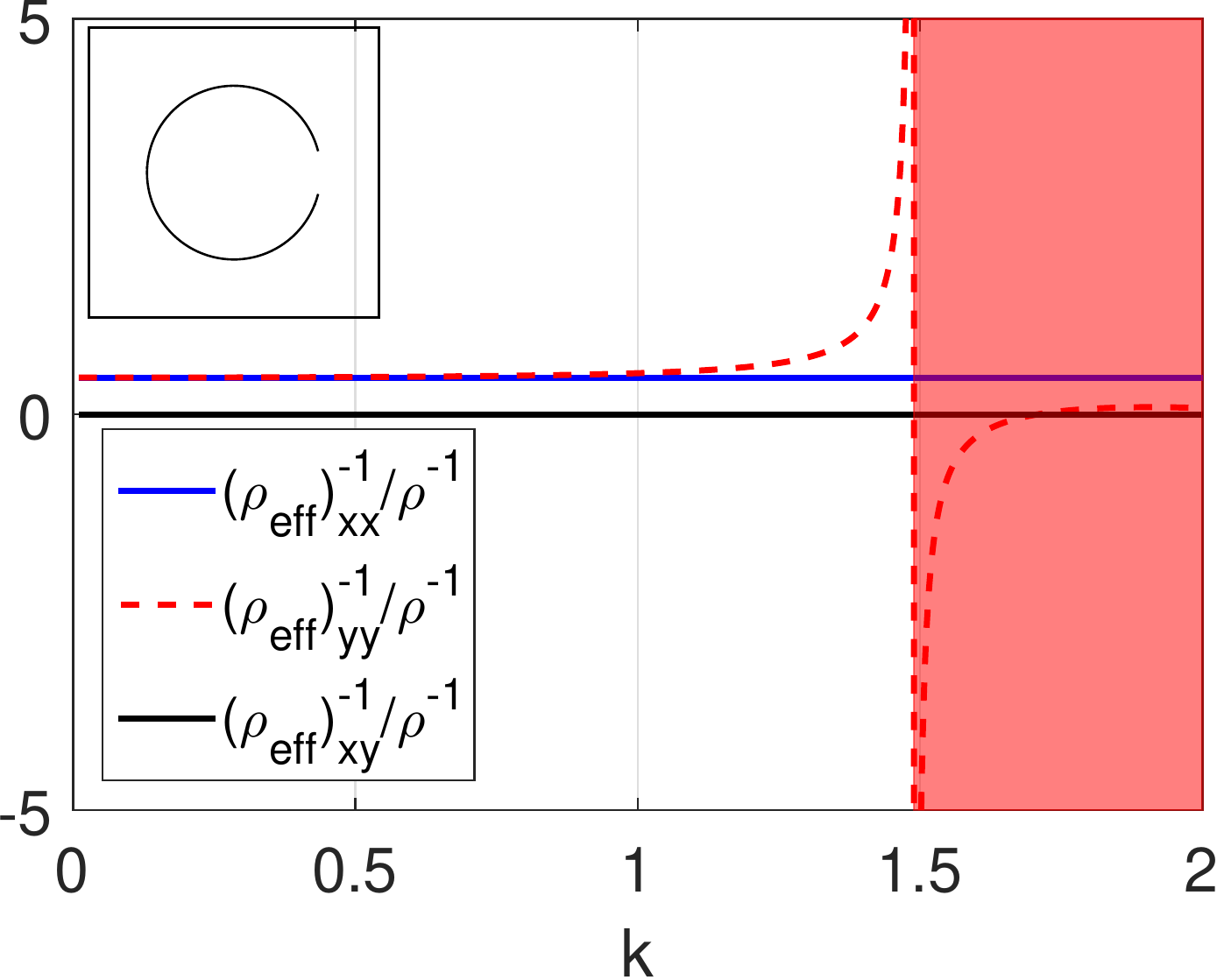}
\label{fig:rhoinvanisothin}}
\subfloat[Subfigure 3 list of figures text][]{
\includegraphics[width=0.425\textwidth]{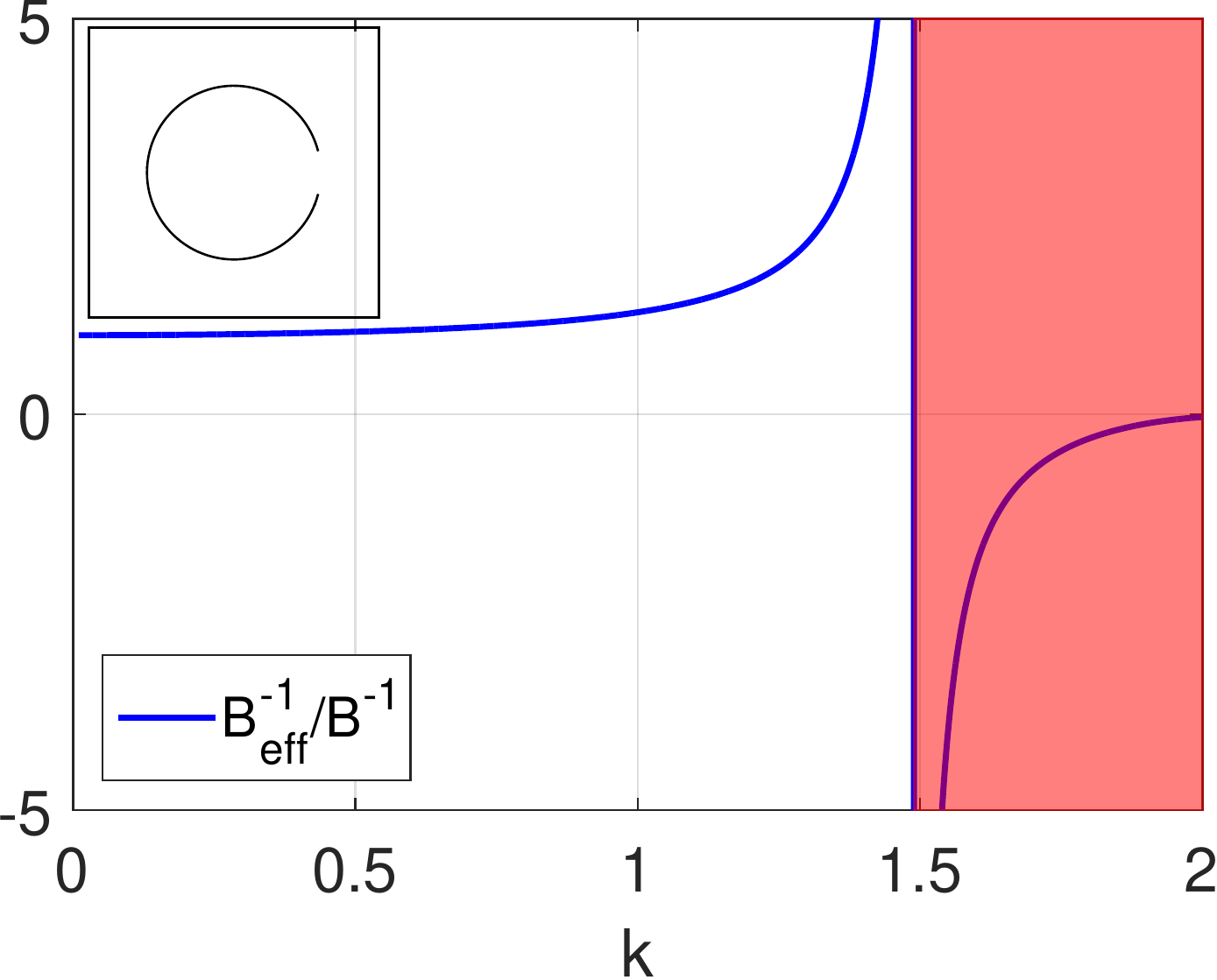}
\label{fig:Binvanisothin}} \\
\subfloat[Subfigure 2 list of figures text][]{
\includegraphics[width=0.425\textwidth]{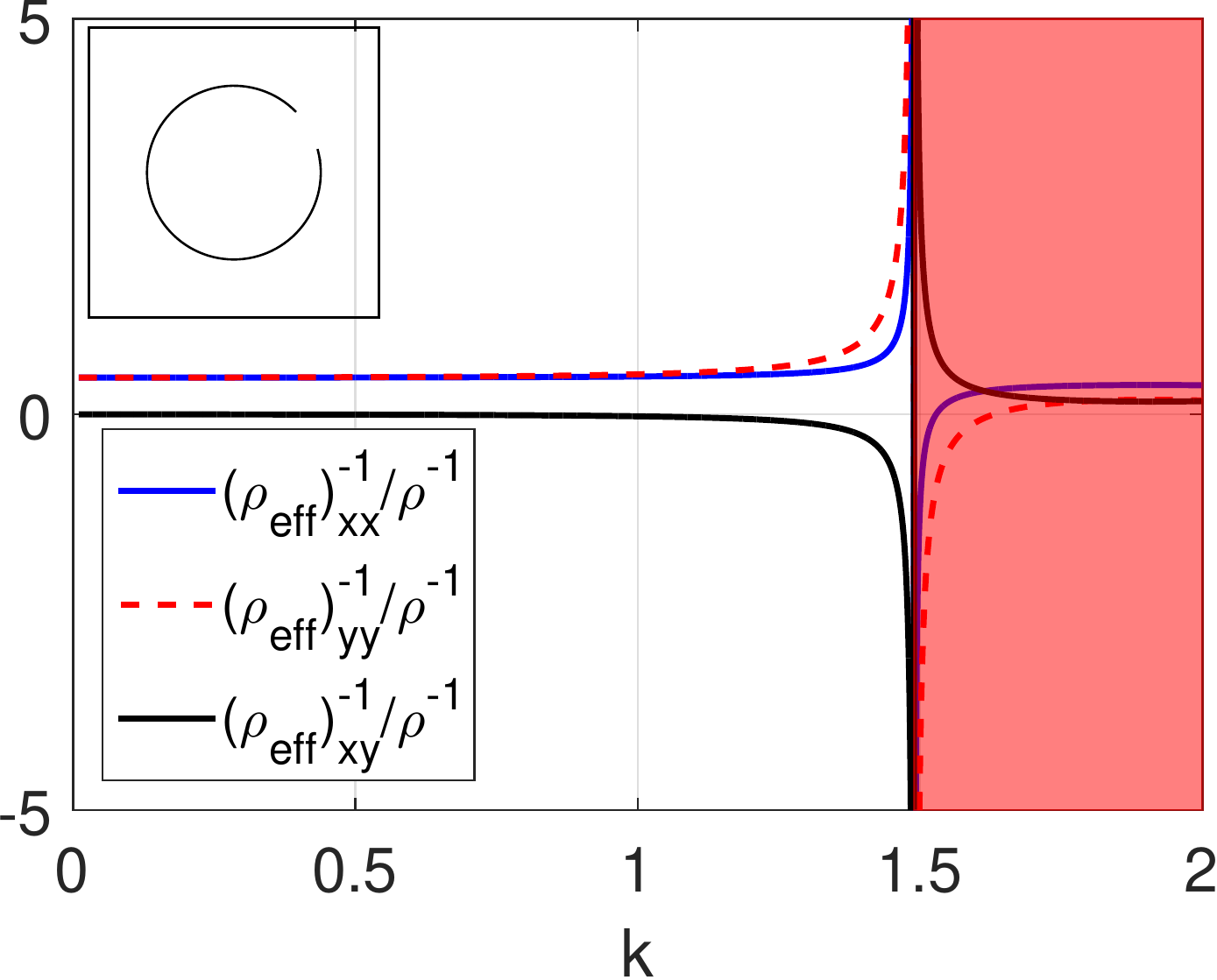}
\label{fig:rhoinvanisothinrot}}
\subfloat[Subfigure 3 list of figures text][]{
\includegraphics[width=0.425\textwidth]{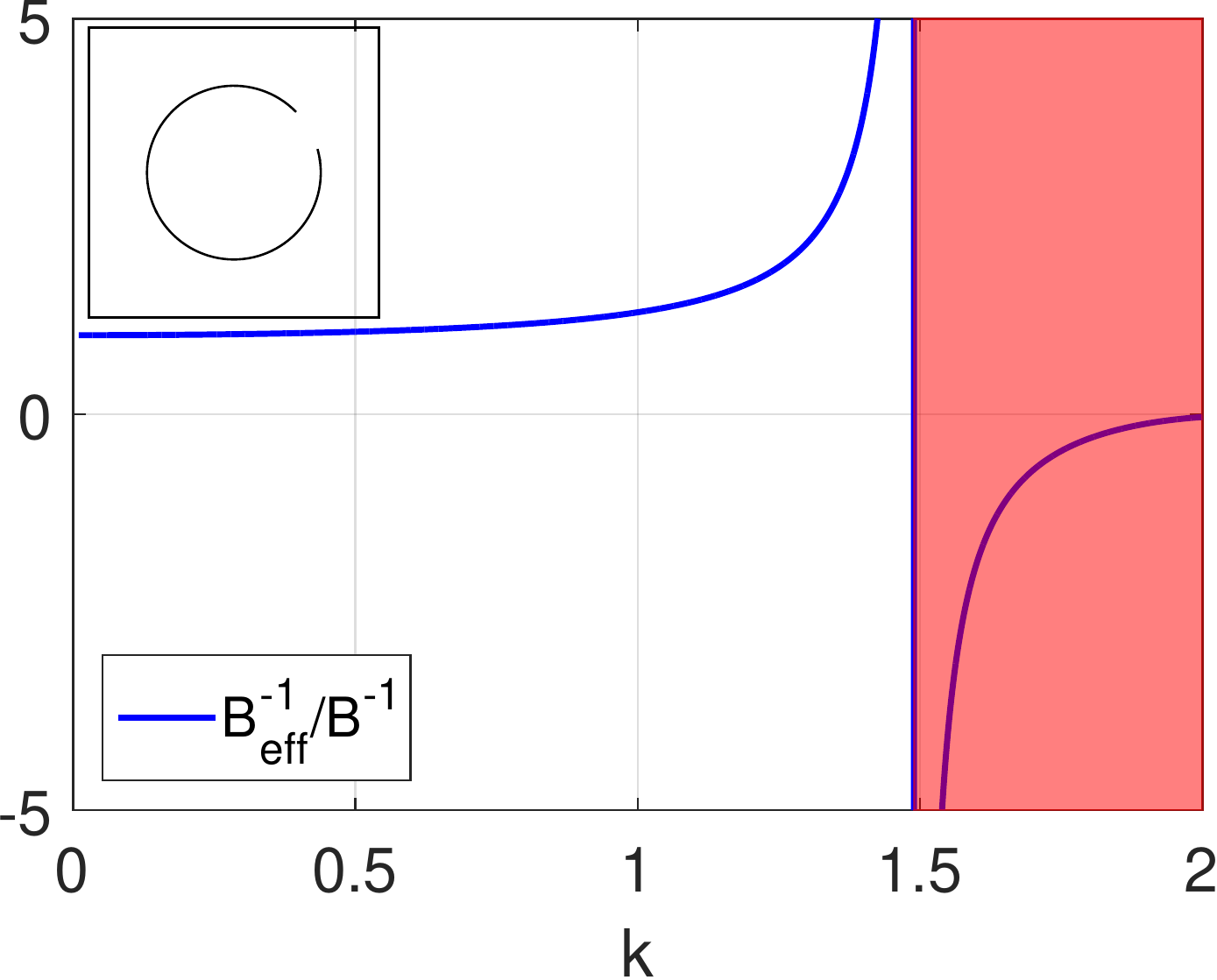}
\label{fig:Binvanisothinrot}}
\caption{The effective inverse density and effective inverse Bulk modulus   for a two-dimensional array of  thin-walled Helmholtz resonators \protect\subref{fig:rhoinvisothin},\protect\subref{fig:Binvisothin} within an isotropic approximation \eqref{eq:combo2} for $\theta_0 = 0$,    \protect\subref{fig:rhoinvanisothin},\protect\subref{fig:Binvanisothin} within an anisotropic approximation \eqref{eq:anisorhoBrespfuncs} for $\theta_0 = 0$, and  \protect\subref{fig:rhoinvanisothinrot},\protect\subref{fig:Binvanisothinrot} within an anisotropic approximation \eqref{eq:anisorhoBrespfuncs} for $\theta_0 = \pi/6$.  The shaded red regions denote  all $k$-values above the Helmholtz resonance/cut-off frequency given by $1-2f_\varepsilon \approx 0$ and \eqref{eq:helmresfreqaniso}, respectively.  Inset: corresponding fundamental unit cells. In the above figures we use $\bar{d} = 1$,   $\theta_\mathrm{ap} = \pi/12$,  and $\bar{b} = 0.3$.}
\label{fig:respfuncsthinwall}
\end{figure}

\begin{figure}[t]
\centering
\subfloat[Subfigure 6 list of figures text][]{
\includegraphics[width=0.425\textwidth]{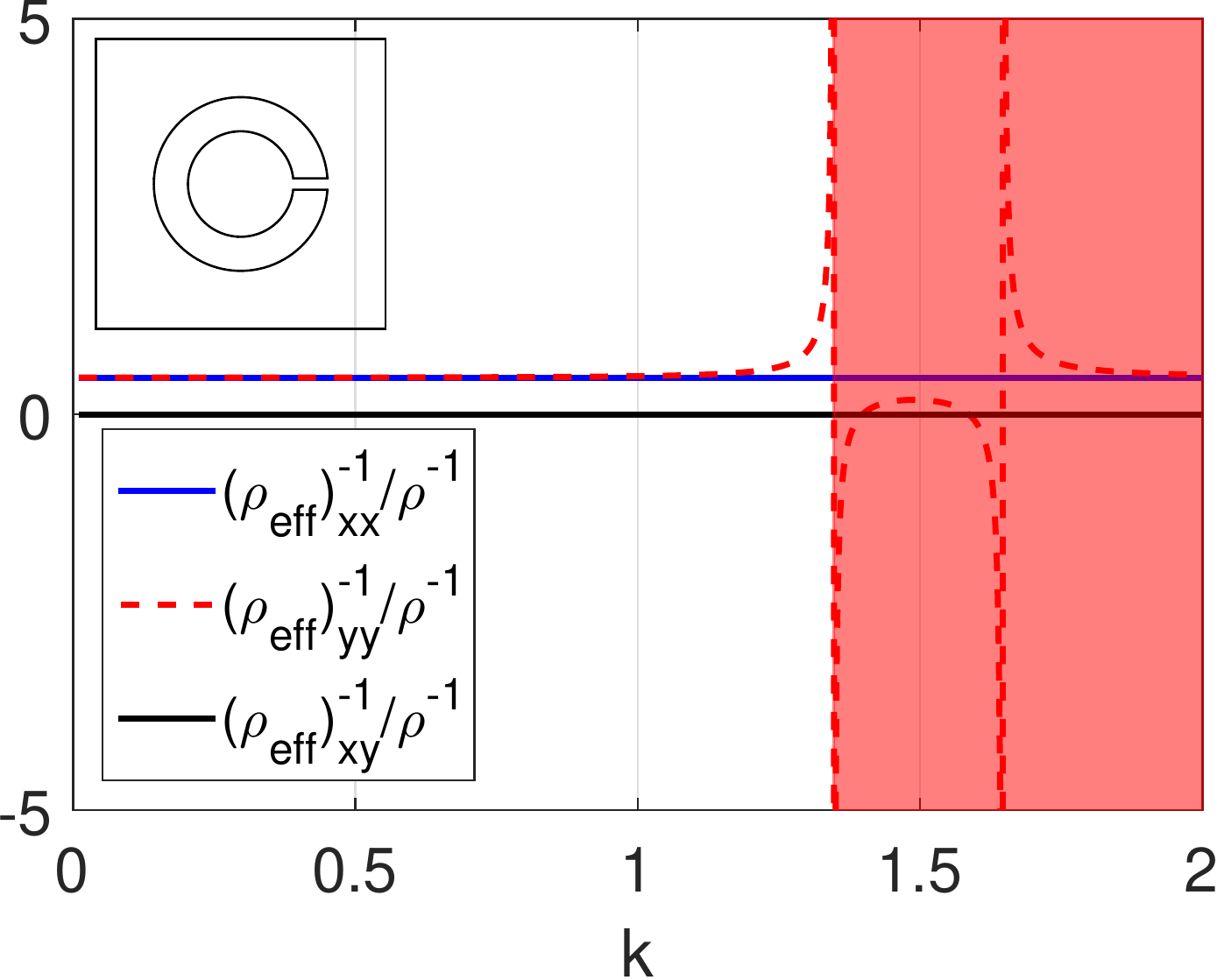}
\label{fig:rhoinvisothin2}}
\subfloat[Subfigure 1 list of figures text][]{
\includegraphics[width=0.425\textwidth]{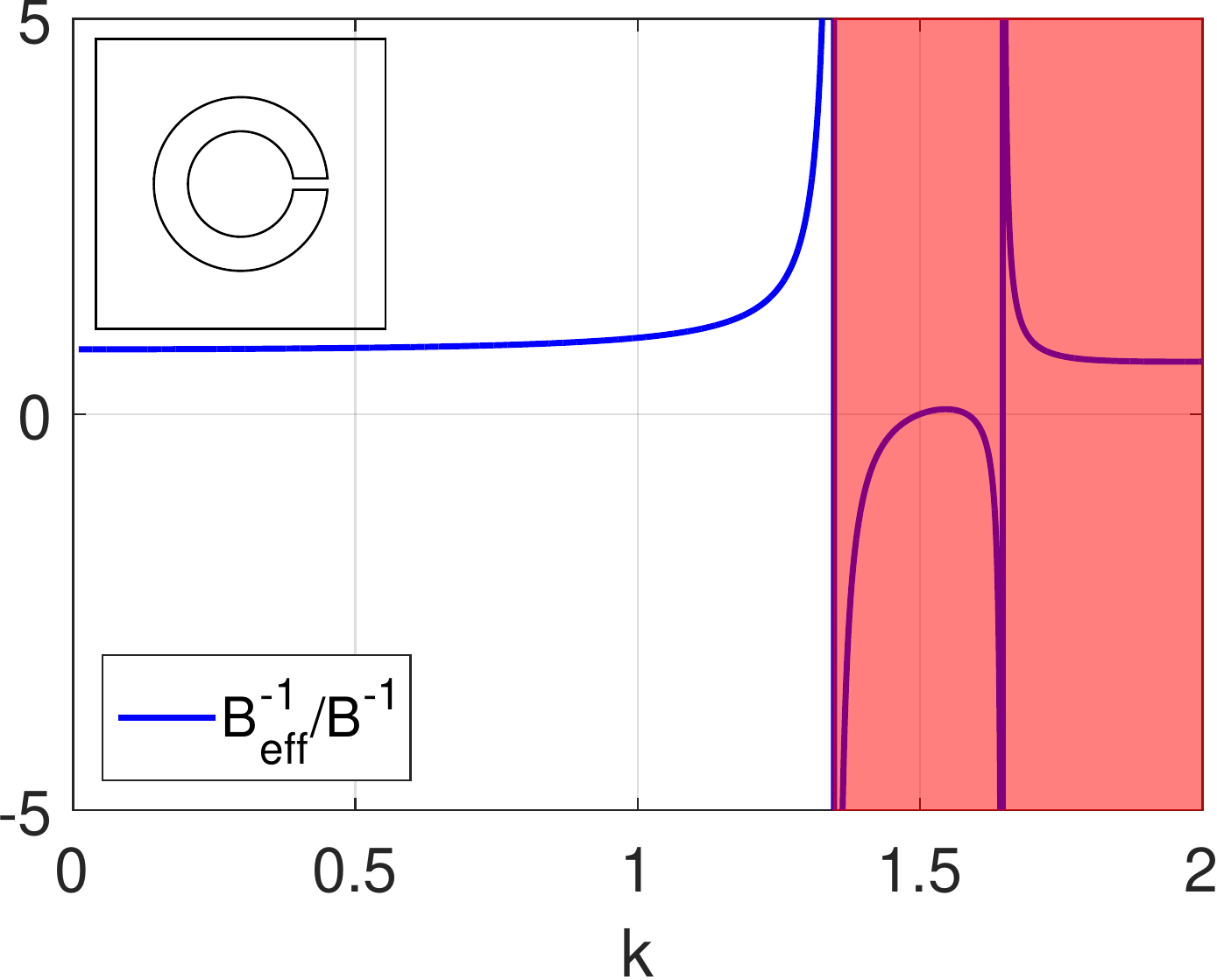}
\label{fig:Binvisothin2}}\\
\subfloat[Subfigure 2 list of figures text][]{
\includegraphics[width=0.425\textwidth]{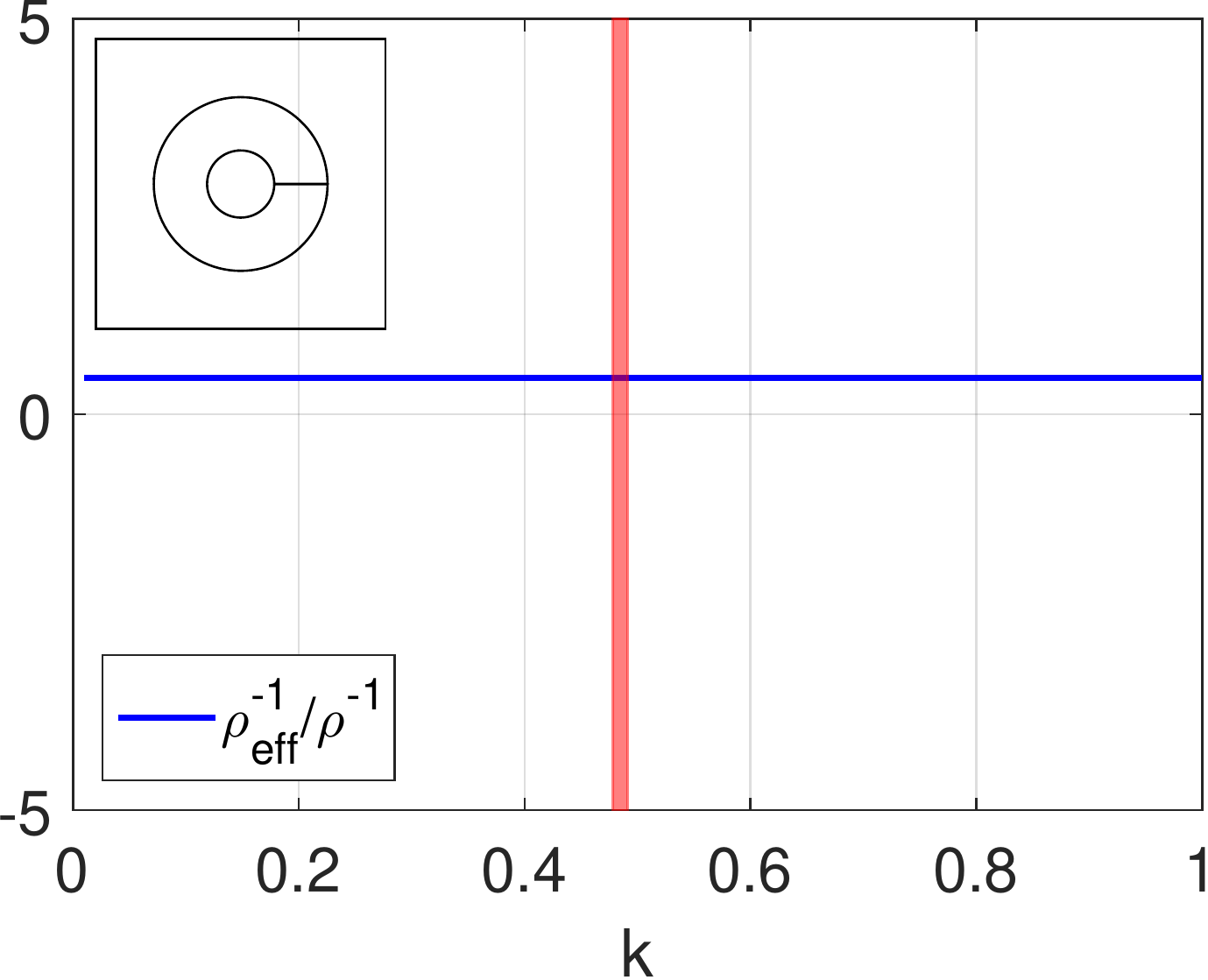}
\label{fig:rhoinvanisothin2}}
\subfloat[Subfigure 3 list of figures text][]{
\includegraphics[width=0.425\textwidth]{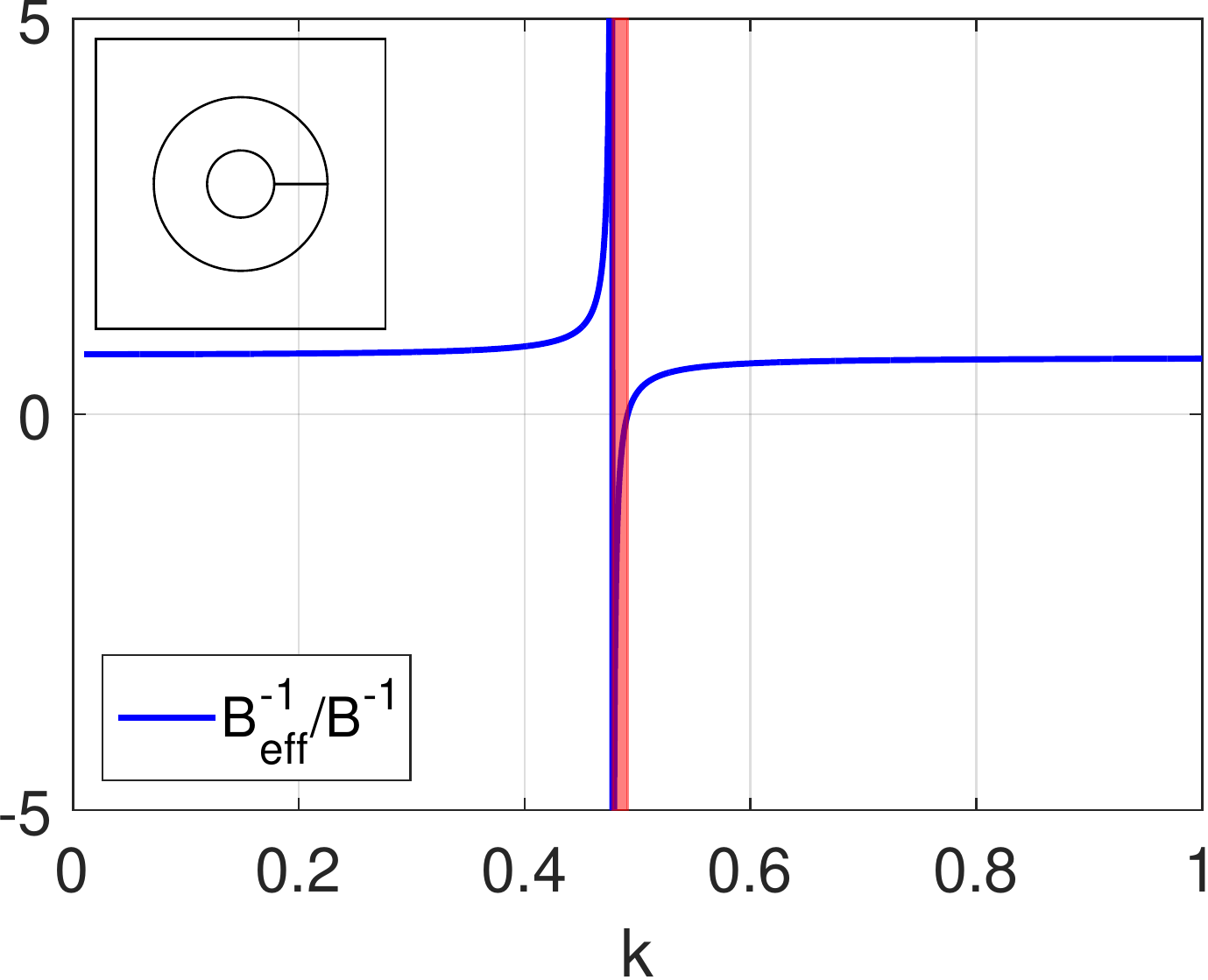}
\label{fig:Binvanisothin2}}
\caption{The effective inverse density and effective inverse Bulk modulus     \protect\subref{fig:rhoinvisothin2},\protect\subref{fig:Binvisothin2} for a two-dimensional array of  moderately thick-walled Helmholtz resonators  within an anisotropic approximation \eqref{eq:anisorhoBrespfuncs}; \protect\subref{fig:rhoinvanisothin2},\protect\subref{fig:Binvanisothin2}   for an array of {\it specially-scaled}   Helmholtz resonators     within an isotropic approximation \eqref{eq:combo2}.  Inset: corresponding fundamental unit cells.  In  \protect\subref{fig:rhoinvisothin2},\protect\subref{fig:Binvisothin2} the shaded red region denotes all $k$-values above the Helmholtz resonance/cut-off frequency  \eqref{eq:helmresfreqaniso}, where we use $\bar{d} = 1$, $\theta_0 = 0$, $\theta_\mathrm{ap} = \pi/48$, $h=3$, and $\bar{b} = 0.3$. In \protect\subref{fig:rhoinvanisothin2},\protect\subref{fig:Binvanisothin2}  the shaded red region denotes the width of the first band gap as shown in Figure \ref{fig:varyingh}\protect\subref{fig:thickmultipolesubfig3}, where we use $\bar{d} = 1$, $\theta_0 = 0$, $\theta_\mathrm{ap} = \pi/1024$, $h=100$, and $\bar{b} = 0.3$.}
\label{fig:respfuncsthickerwalls}
\end{figure}

In Figure \ref{fig:convergenceboth} we examine  the band diagram for two {\it specially-scaled} resonator configurations, comparing results for the regularised system \eqref{eq:dispeqsystemgnptfinal} under different truncations against those obtained using finite element methods. In general, we find that a dipole approximation (dashed black lines)   works quite well up to the saddle point frequency of the second band surface, with quadrupole corrections  required only for higher frequencies. Hence, in  the figures that follow (Figures \ref{fig:varyingh}--\ref{fig:varyingf}) we consider a quadrupole system truncation ($L=3$)  for overall accuracy and a dipole system truncations to derive asymptotic descriptions. A significant change is observed in the Helmholtz resonance frequency (equivalently, the maximum frequency of the first band surface) between the two configurations in Figure \ref{fig:convergenceboth},  which is attributable to the change in aperture angle. The impact of this parameter is discussed in further detail below in Figure \ref{fig:trackhelmres}.

In Figure \ref{fig:varyingh} we compute the band diagrams for a {\it specially-scaled} Helmholtz resonator array  as the channel aspect ratio $h$ is varied. Here we observe that the multipole system, within a quadrupole approximation, is likewise able to recover the spectral behaviour to excellent accuracy and that the isotropic approximation \eqref{eq:dispeqdipoleleadinginpt2} is able to recover the first band, determine the width of the first band gap, and   describe   the second band surface at low frequencies, provided that $h$ is not too large. For all values of $h$ we find that the isotropic approximation works surprisingly well over the range of the first band surface.

Similarly, Figure \ref{fig:varyingthetaap} examines the impact of varying the aperture half-angle $\theta_\mathrm{ap}$ on the band diagram for   {\it specially-scaled}  resonator arrays. Once more, we find that the multipole system exhibits excellent performance, and that the isotropic  approximation \eqref{eq:dispeqdipoleleadinginpt2} also performs surprisingly well, provided that the aperture half-angle is not too large. In fact, for $\theta_\mathrm{ap} = \pi/16$ we observe that the first and second band surfaces are degenerate along $\rmM \rmY$. It is clear that the Helmholtz resonance/cut-off frequency is considerably sensitive to varying aperture angle, which we discuss further in        Figure \ref{fig:trackhelmres} below.

In Figure \ref{fig:varyingf} we determine the band diagrams for a {\it specially-scaled} Helmholtz resonator array, as the filling fraction is varied. As in the preceding figures, the multipole treatment works remarkably well, even as the outer resonator wall  almost touches the boundaries of the unit cell. This is quite surprising given that an ever increasing number of multipole orders are required to compute the band diagram for ideal cylinders in the same limit  \cite{movchan2002asymptotic}. As before, the isotropic approximation \eqref{eq:dispeqdipoleleadinginpt2}  is able to recover the first band surface to exceptional accuracy, and is able to extend into the second band surface for moderate filling fractions. Even as the filling fraction approaches the wall-touching limit, the approximation \eqref{eq:dispeqdipoleleadinginpt2} is still able to determine the width of the first band gap to suitable accuracy.

In Figure \ref{fig:respfuncsthinwall} we examine  the effective inverse density and effective inverse Bulk modulus for an array of thin-walled Helmholtz resonators, specifically comparing the isotropic and anisotropic expressions  \eqref{eq:combo2} and \eqref{eq:anisorhoBrespfuncs}. In Figures  \ref{fig:respfuncsthinwall}\protect\subref{fig:rhoinvisothin},\protect\subref{fig:Binvisothin} we observe that within an isotropic approximation,    the band edge coincides with  the Helmholtz resonance frequency ($2\dbarfeps-1\approx 0$), where the compressibility likewise diverges. As expected from the representations in \eqref{eq:combo2}, frequency dependence is observed in the inverse Bulk modulus alone. In Figures \ref{fig:respfuncsthinwall}\protect\subref{fig:rhoinvanisothin},\protect\subref{fig:Binvanisothin}  we observe an expected frequency dependence in one component  of the effective inverse density matrix as well as for the inverse Bulk modulus, when   anisotropy is considered, see \eqref{eq:anisorhoBrespfuncs}. Here,   $(\rho_\mathrm{eff})^{-1}_{\rmy\rmy}$ also  diverges at the Helmholtz resonance frequency \eqref{eq:helmresfreqaniso}. Similarly, in Figures \ref{fig:respfuncsthinwall}\protect\subref{fig:rhoinvanisothinrot},\protect\subref{fig:Binvanisothinrot}, we observe frequency dependence and divergence at the resonance frequency \eqref{eq:helmresfreqaniso} for all entries in $(\rho_\mathrm{eff})^{-1}_{ij}$ as well as for the inverse Bulk modulus, when $\theta_0 = \pi/6$. By rotating the resonator  we have reduced the symmetry properties of the medium, allowing for stronger anisotropy and dispersion.  In any case, despite the emergence of anisotropy in the effective inverse density, no changes in sign are observed in the coefficients over the range of the first band surface ( $(\rho_\mathrm{eff})^{-1}_{\rmx\rmy}$ is either zero or negative). This suggests that exotic effects such as negative refraction are not supported on the first band surface for Helmholtz resonator arrays, however we expect that negative refraction is supported at higher frequencies where the correct band curvature is exhibited \cite{smith2012negative}.

In Figures \ref{fig:respfuncsthickerwalls}\protect\subref{fig:rhoinvisothin2},\protect\subref{fig:Binvisothin2} we  consider the effective inverse density and inverse Bulk modulus functions, within an anisotropic description, corresponding to a moderately thick-walled resonator array. These  curves are obtained   via \eqref{eq:anisorhoBrespfuncs} with the replacement $f_\varepsilon \mapsto \check{f}_\varepsilon$ (results for the isotropic descriptions are not included here as these are qualitatively similar to those given for the thin-walled case). As  observed in the thin-walled case, we see frequency dependence in both material  tensors  over the frequency range of the first band surface, which is accompanied by unexpected singular behaviour above the Helmholtz resonance frequency. No firm conclusions may be drawn from this, however, as this behaviour lies outside the region of validity for the expressions. In Figures  \ref{fig:respfuncsthickerwalls}\protect\subref{fig:rhoinvanisothin2},\protect\subref{fig:Binvanisothin2} we consider  a {\it specially-scaled} resonator array  (i.e., \eqref{eq:combo2} with the replacement $f_\varepsilon \mapsto \dbarfeps$) which exhibits much more interesting behaviour. In particular, we observe a pole in $B_\mathrm{eff}^{-1}$ at low frequencies, where the width of the first band gap may be defined as the interval between the Helmholtz resonance frequency given by $1-2\dbarfeps \approx 0$ and the zero of $B_\mathrm{eff}^{-1}$; that is, the first band gap corresponds to the  interval where the effective bulk modulus is negative. Once this physical parameter returns to positive values we find that these isotropic descriptions extend well into the range of the second band surface, as seen in Figure \ref{fig:varyingh}\protect\subref{fig:thickmultipolesubfig3}. For reference,  we estimate our descriptions \eqref{eq:combo2} to hold over the approximate range $0\leq k \leq 1.5$ for this example, which supports the assertion that  the isotropic description for {\it specially-scaled} resonator arrays appears to be  valid over a significantly broader frequency range than for the resonators discussed in Part I.

Finally, in Figure \ref{fig:trackhelmres}, to complement the band diagram figures outlined in Figures \ref{fig:varyingh} and \ref{fig:varyingthetaap}, we solve $2\dbarfeps - 1\approx 0$ to track the Helmholtz resonance/cut-off frequency as the aspect ratio $h$ and aperture half-angle $\theta_\mathrm{ap}$ are varied. In Figure \ref{fig:helmtrack1} we observe that the cut-off frequency for the first band surface  is able to achieve a minimum of $k \approx 0.369$ at $h\approx 52$ for a fixed $\theta_\mathrm{ap} = \pi/1024$, and that as the aspect ratio $h$ becomes large, the cut-off frequency grows  larger with polynomial scaling.  The possibility of achieving a cutoff frequency minimum, by tuning the wall thickness, may prove useful for those involved in the design of acoustic metamaterials. Likewise,   Figure \ref{fig:helmtrack2} demonstrates that the  resonant frequency decreases with decreasing aperture half-angle, as expected, and   scales as $k_\mathrm{max} \propto \theta_\mathrm{ap}^{1/2}$ as $\theta_\mathrm{ap}\rightarrow 0$.

\begin{figure}[t]
\centering
\subfloat[Subfigure 6 list of figures text][]{
\includegraphics[width=0.4275\textwidth]{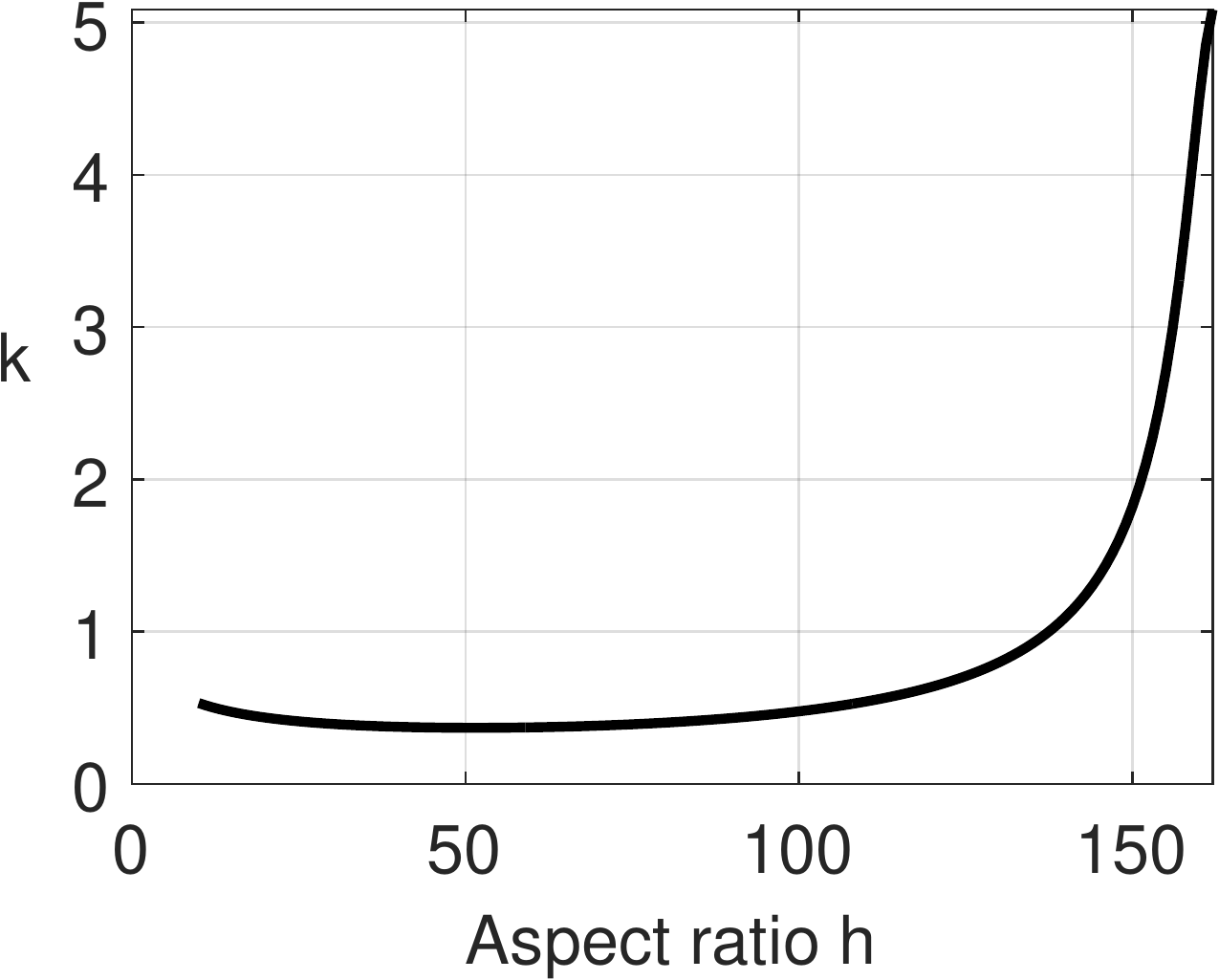}
\label{fig:helmtrack1}}
\subfloat[Subfigure 1 list of figures text][]{
\includegraphics[width=0.4275\textwidth]{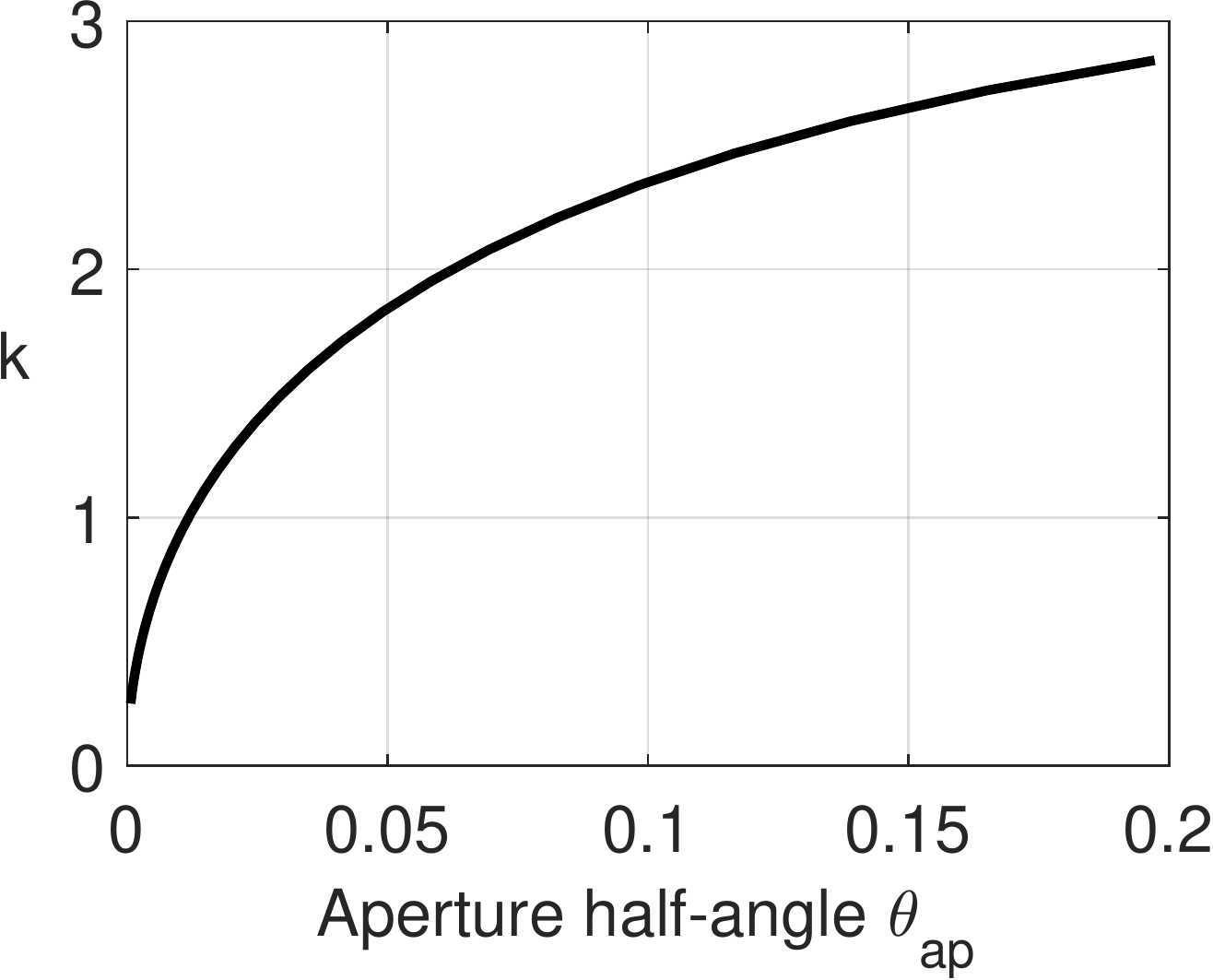}
\label{fig:helmtrack2}}

\caption{Tracking the cut-off/Helmholtz resonance frequency (by solving $2\dbarfeps-1\approx 0$) in the specially-scaled  setting   as \protect\subref{fig:helmtrack1} the   aspect ratio of the resonator neck $h =  {\emm} /  {\ell}$   is varied, as shown in Figure \ref{fig:varyingh}, and as \protect\subref{fig:helmtrack2}    the aperture half-angle $\theta_\mathrm{ap}$ is varied, as demonstrated in Figure \ref{fig:varyingthetaap}.
 \label{fig:trackhelmres}}
  \end{figure}

\section{Discussion} \label{sec:discus}
In this paper, we have presented a matched asymptotic-multipole procedure   for determining the band structure of an acoustic metamaterial comprising a two-dimensional array  of  Helmholtz resonators that possess   large  wall thicknesses        and     narrow neck widths. We have also derived a compact dispersion equation which is able to describe the first band surface, first band gap, and frequencies well into the second band surface, over a range of practical settings. In addition, we have outlined a homogenisation procedure for thin-walled, moderately thick-walled, and {\it specially-scaled} resonator arrays, presenting analytical forms for the effective inverse density tensor and the effective inverse Bulk modulus. The  effective response functions   derived extend well beyond the quasistatic limit to higher frequencies, depending on the resonator geometry. Furthermore, we have also derived closed-form representations for the Helmholtz resonance frequency in the {\it specially-scaled} setting.  

We demonstrate that the {\it specially-scaled} resonators are able to achieve extremely low Helmholtz resonance frequencies, in contrast to   thin- and moderately-thick walled resonators presented in Part I, which has a marked impact on the performance of the array. By incorporating long neck widths $2 \bar{\emm}$ into the formulation we provide an additional degree of freedom for controlling the frequency range of the first band surface,   indirectly controlling features such as the low-frequency phase and group velocity. Incidentally, we remark that   the Helmholtz resonance frequency is often approximated   in the literature via the form  \cite[Eq. (5.3.12)]{howe1998acoustics}  
\begin{equation} 
\label{eq:helmholtzcommonaprox}
\omega_\mathrm{max} \approx \sqrt{\frac{B}{\rho}} \sqrt{\frac{A}{LV}}, \quad \mbox{ or equivalently}, \quad k_\mathrm{max} \approx \frac{1}{\bar{a}}\sqrt{ \dfrac{1}{\pi \bar{\emm}/\bar{\ell}}}
\end{equation}
   where $A$ is the total aperture width $2\bar{\ell}$, $L$ denotes the length of the resonator neck $2\bar{\emm}$, and $V$ denotes the enclosed   resonator area $\pi \bar{a}^2$. The major disadvantage of \eqref{eq:helmholtzcommonaprox} is that the  approximation   is  quite crude, as it treats the neck as distinct from the enclosed volume, and as a result frequently requires correction factors and an effective neck length  to recover accuracy. We stress that our Helmholtz resonance expressions derived   in \eqref{eq:rescondspecscaled1}, \eqref{eq:rescondspecscaled2}, and in Part I, do not require any such corrections, and apply to both single resonators and arrays of resonators alike, based purely on asymptotic scale separation via the small parameter $\varepsilon$.

On comparing results from Parts I and II, we find that anisotropy in both the band structure and in the effective tensors is greatest for thin-walled resonators, with anisotropic effects considerably reduced as we approach the {\it specially-scaled} limit.  In fact, we find that {\it specially-scaled} resonators may be treated as {\it almost-isotropic} media. Note however that   results from Part I are not   recovered as $\bar{\emm}\rightarrow 0$ within the current formulation, or vice versa, due to the different underlying assumptions of the inner solutions; hence the reason for treating them separately. A  core advantage of our matched asymptotics-multipole treatment is that it avoids the need for extensive fully-numerical procedures, such a finite-element methods, for determining the low-frequency band structure of acoustic metamaterials. Fully-numerical procedures require intensive meshing/sampling inside the neck region as  it   becomes increasingly thin, and although the computational domain is   two-dimensional, meshing requirements can massively increase  computation times and resource requirements.  In contrast, our asymptotic dispersion equations provide rapidly-evaluable closed-form representations for band surfaces over a wide frequency range.   It is of interest to apply the techniques outlined here to related geometries,  such as resonators with long maze-like channels \cite{quan2018maximum} or resonators nested within resonators \cite{elford2011matryoshka,montiel2020planar}, and this is presently under investigation by the authors.

In relation to our homogenisation treatment, we emphasise that composite materials and metamaterials  have the potential to   exhibit   generalised    constitutive relations \cite{milton2007modifications,willis2011effective}. Such behaviour  occurs  widely across the acoustics and elasticity literature, where these materials are known as  {\it Willis media}, as well as in the electromagnetics literature, where they are known as {\it bi-isotropic} or {\it bi-anisotropic media}     \cite{lindell1994electromagnetic}.   A significant body of literature has developed in recent years with such effects in mind  \cite{norris2012analytical,torrent2015resonant,muhlestein2016reciprocity,muhlestein2017experimental}.  It would be of interest to  examine Willis coupling effects in resonator arrays in further detail. Recently, a body of experimental and theoretical work has emerged demonstrating   analogues to Willis coupling in   wave scattering by a single thick-walled Helmholtz resonator \cite{quan2018maximum,melnikov2019acoustic} which could offer interesting potential applications.

Finally,  there are points regarding   passivity and causality in resonator arrays, and the behaviour of   matched asymptotic expansion solutions, which are worth investigating and this will be reported on by the authors in a forthcoming article.

 \appendix

 \section{Asymptotic forms of the lattice sums} \label{sec:asymptoticsSmY}

As discussed in the Appendix of Part I, the lattice sums $S_m^\rmY$ emerge frequently in the study of periodic media, but  are conditionally convergent in their most direct  form and so must   be  regularised in order to retrieve physically meaningful results. 
The convergent expressions given in Part I are used in all relevant numerical computations, whereas for the purposes of asymptotic analysis we follow  the procedures outlined in   \cite{mcphedran1996low,chen2018evaluation}, to derive  and present  leading terms for the  first few lattice sums of a square array     in the form:
\begin{equation}
\label{eq:lattsum1stordertermsmany}
\begin{aligned}
S_0^\rmY &\sim-\frac{4 f}{\pi  b^2 \left(k_\rmB^2-1\right)} + \frac{1}{\pi}\left[ -2 \gamma_\rme + \log\left(\frac{f \Gamma(\tfrac{1}{4})^4}{\pi^2 b^2}\right) \right],
\\
S_1^\rmY &\sim -\frac{4 \rmi f k_\rmB}{\pi  b^2 \left(k_\rmB^2-1\right)}  \rme^{\rmi \theta_\rmB} + \left(\frac{\rmi k_\rmB}{\pi} \right) \rme^{\rmi \theta_\rmB},
\\
S_2^\rmY &\sim
\frac{4 f k_\rmB^2}{\pi  b^2 \left(k_\rmB^2-1\right)}  \rme^{2 \rmi \theta_\rmB} + \left[  \frac{k_\rmB^2  \Gamma  (\frac{1}{4} )^8}{384\pi^5}\rme^{-2\rmi\theta_\rmB}  -\frac{k_\rmB^2}{2\pi}\rme^{2\rmi\theta_\rmB}\right],
\\
S_3^\rmY &\sim
\frac{1}{b^2}\left[
\frac{4 \rmi f k_\rmB^3 }{\pi     \left(k_\rmB^2-1\right)} \rme^{3 \rmi \theta_\rmB} -\frac{\rmi    k_\rmB f\Gamma  (\tfrac{1}{4} )^8 }{48\pi ^5  } \rme^{-\rmi \theta_\rmB} \right], 
\\
S_4^\rmY &\sim
-\frac{f^2 \Gamma  (\frac{1}{4} )^8}{10 \pi ^5 b^4},
\end{aligned}
\qquad \hspace{-30mm}
\begin{aligned}
S_5^\rmY &\sim
-\frac{2\rmi f^2 k_\rmB  \Gamma  (\tfrac{1}{4} )^8}{5 \pi^5 b^4}\rme^{\rmi \theta_\rmB},
\\
S_6^\rmY\sim
\frac{1}{b^4}\left[
\frac{f^2 k_\rmB^2 \Gamma  (\tfrac{1}{4} )^8}{\pi^5  } \rme^{2\rmi\theta_\rmB} \right. &+ \left. \frac{5 f^2 k_\rmB^2 \Gamma  (\tfrac{1}{4} )^{16}}{960\pi^9 } \rme^{-2\rmi\theta_\rmB}\right], \\
S_7^\rmY&\sim
-\frac{\rmi f^3 k_\rmB \Gamma  (\tfrac{1}{4} )^{16} }{14 \pi^9 b^6}\rme^{-\rmi \theta_\rmB},
\\
S_8^\rmY&\sim
-\frac{3 f^4 \Gamma  (\tfrac{1}{4} )^{16} }{5\pi^9 b^8},
\\
S_9^\rmY&\sim
-\frac{24 \rmi \rme^{\rmi \theta_\rmB} f^4 k_\rmB \Gamma  (\tfrac{1}{4} )^{16} }{5 \pi^9 b^8}.
\end{aligned}
\end{equation}
 For reference, we remark that these sums are strictly  defined in terms of the lattice period $d$. However, we have introduced  $f = \pi b^2/d^2$, the cylinder area fraction, here since the limit of $b \rightarrow 0$ (with $f$ fixed) is considered in  our analysis.

\section*{Acknowledgements}
   I.D.A. acknowledges  support from a Royal Society Industry Fellowship. This work was also supported by EPSRC grant no EP/R014604/1 whilst I.D.A. held the position of Director of the Isaac Newton Institute Cambridge.

\bibliographystyle{RS.bst}

\end{document}